\documentclass[format=acmsmall, review=false, screen=true]{acmart}

\usepackage[T1]{fontenc}
\graphicspath{{images/}}
\usepackage{pifont}
\usepackage{todonotes} 
\usepackage{url} 
\usepackage{color}
\usepackage{enumerate} 
\usepackage{array}
\usepackage{multirow}
\usepackage{float}
\usepackage{enumitem}
\usepackage{amsmath}
\usepackage{verbatim}
\usepackage{footnote}
\usepackage{adjustbox}
\usepackage{multirow}
\usepackage{array}
\usepackage{makecell}

\usepackage{xcolor}

\newcounter{tab_footnote_cnt}
\newcommand*{\tabindent}{\hspace*{0.1cm}}

\usepackage{booktabs} 

\acmJournal{CSUR}

\setcopyright{acmlicensed}

\begin{document}
\title[A Survey on Malicious Domains Detection]{A Survey on Malicious Domains Detection through\\DNS Data Analysis}  

\author{Yury Zhauniarovich}
\orcid{0000-0001-9116-0728}
\affiliation{%
  \institution{Qatar Computing Research Institute, HBKU}
  \city{Doha}
  \country{Qatar}
}
\email{yzhauniarovich@hbku.edu.qa}

\author{Issa Khalil}
\affiliation{%
  \institution{Qatar Computing Research Institute, HBKU}
  \city{Doha}
  \country{Qatar}
}
\email{ikhalil@hbku.edu.qa}

\author{Ting Yu}
\affiliation{%
  \institution{Qatar Computing Research Institute, HBKU}
  \city{Doha}
  \country{Qatar}
}
\email{tyu@hbku.edu.qa}

\author{Marc Dacier}
\affiliation{%
  \institution{Eurecom}
  \city{Sophia Antipolis}
  \country{France}
}
\email{dacier@eurecom.fr}

\begin{abstract}
Malicious domains are one of the major resources required for adversaries to run attacks over the Internet. Due to the important role of the Domain Name System (DNS), extensive research has been conducted to identify malicious domains based on their unique behavior reflected in different phases of the life cycle of DNS queries and responses. Existing approaches differ significantly in terms of intuitions, data analysis methods as well as evaluation methodologies. This warrants a thorough systematization of the approaches and a careful review of the advantages and limitations of every group.

In this paper, we perform such an analysis. In order to achieve this goal, we present the necessary background knowledge on DNS and malicious activities leveraging DNS. We describe a general framework of malicious domain detection techniques using DNS data. Applying this framework, we categorize existing approaches using several orthogonal viewpoints, namely (1) sources of DNS data and their enrichment, (2) data analysis methods, and (3) evaluation strategies and metrics. In each aspect, we discuss the important challenges that the research community should address in order to fully realize the power of DNS data analysis to fight against attacks leveraging malicious domains.


\end{abstract}

%
%

%
%




\maketitle

\renewcommand{\shortauthors}{Y. Zhauniarovich et al.}

\section{Introduction}
\label{sec:introduction}
It is well known that the Internet is being used continuously to run attacks against different targets. Benign services and protocols are being misused for various malicious activities: to disseminate malware, to facilitate command and control (C\&C) communications, to send spam messages, to host scam and phishing webpages. Clearly, it is very important to detect the origins of such malvolent activities, be it by identifying an URL, a domain name or an IP address. Many approaches have been proposed for such purpose: analysis of network traffic~\cite{Effort_Shin2012, TrafficAggregationForMalwareDetection_Yen2008}, inspection of web content~\cite{Prophiler_Canali2011, BInspect_Eshete2013}, URL scrutiny~\cite{LearningToDetectMaliciousURLs_Ma2011}, or using a combination of those techniques~\cite{RbSeeker_Hu2009, BeyondBlacklists_Ma2009}. On top of these, one of the most promising directions relies on the analysis of the \emph{Domain Name System} data.

\emph{Domain Name System} (DNS) protocol is an essential part of the Internet. It maps tough-to-remember Internet Protocol (IP) addresses to easy memorable domain names. Detection of malicious domains through the analysis of DNS data has a number of benefits compared to other approaches. First, DNS data constitutes only a small fraction of the overall network traffic, what makes it suitable for analysis even in large scale networks which cover large areas. Moreover, caching, being an integral part of the protocol, naturally facilitates further decrease the amount of data to be analyzed, allowing researchers to analyze even the DNS traffic coming to Top Level Domains~\cite{Kopis_Antonakakis2011}. Second, the DNS traffic contains a significant amount of meaningful features to identify domain names associated to malicious activities. Third, many of these features can further be enriched with associated information, such as AS number, domain owner, etc. providing an even richer space exploitable for detection. The large amount of features and the vast quantity of traffic data available have made DNS traffic a prime candidate for experimentation with various machine learning techniques applied to the context of security. Forth, although the solutions to encrypt DNS data, like DNSCrypt~\cite{DNSCrypt} exist, still a large fraction of DNS traffic remains unencrypted, making it available for the inspection in various Internet vantage points. Last but not least, sometimes researchers are able to reveal attacks at their early stages or even before they happen due to some traces left in the DNS data.  

The purpose of this paper is to survey all the approaches that aim at detecting domains involved in malicious activities through the analysis of DNS data. To do so, we have built a comprehensive bibliography by collecting papers from several sources. First, we have crawled 4 major digital libraries, namely, ACM\footnote{\url{http://dl.acm.org}}, IEEEXplore\footnote{\url{http://ieeexplore.ieee.org}}, Springer\footnote{\url{https://rd.springer.com/}} and Scopus\footnote{\url{https://www.scopus.com/}}, feeding them with a search string consisting of keywords relevant to the area. Second, we asked credible experts to provide us the most pertinent articles. Third, we extracted from these papers the references which were not included so far in our compiled list. Additionally, we continued to monitor major conferences for any relevant new work appearing in the area. Note that the focus of this paper is not limited to domains involved in specific types of malicious activities, as done in~\cite{SurveyOfBotnetAndBotnetDetection_Feily2009, TaxonomyOfBotnetBehaviorDetectionAndDefense_Khattak2014, Alieyan2015, Dhole16}, that provide surveys specifically about botnets; or in~\cite{PhishingDetection_Khonji2013, FeatureSelectionForPhishingDetection_Zuhair2016}, \cite{SurveyOnWebSpamDetection_Spirin2012} and~\cite{MaliciousUrlDetection_Sahoo2017}, that cover areas of phishing, web spam and malicious URLs detection correspondingly.

\begin{figure}[t!]
    \centering
    \includegraphics[width=0.75\textwidth]{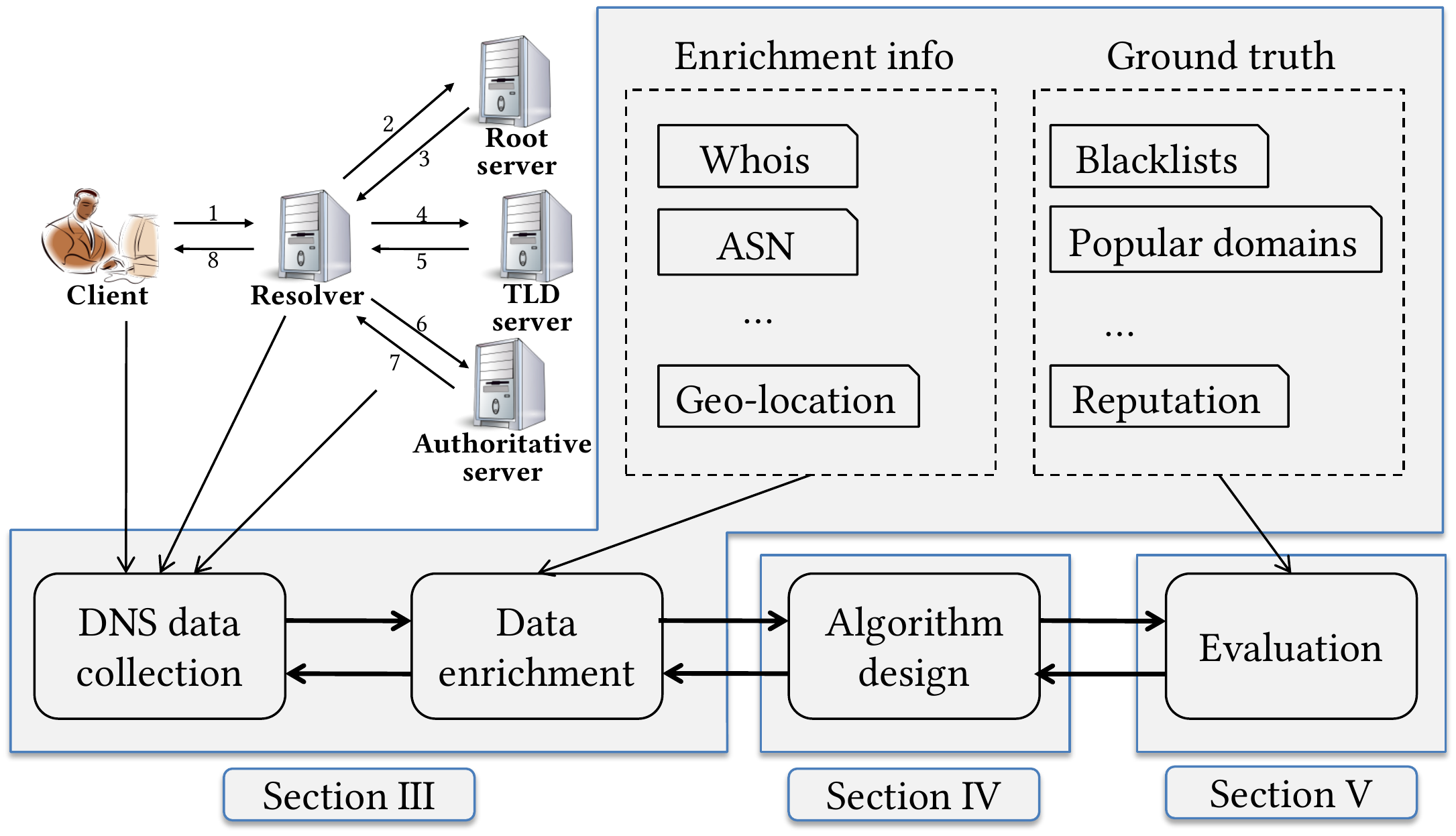}
    \caption{The general process to design a DNS data-based technique to detect malicious domains.}
    \label{fig:process}
\end{figure}

We have carefully read each paper in our study list and extracted the information that could help us to cover the targeted research topic. The first observation we have made, is that this research area is relatively new. The seminal paper~\cite{PassiveDnsReplication_Weimer2005}, which led to the area as we know it today, dates back to 2005. Authored by Florian Weimer, it was the very first published paper not only to consider using DNS records to detect malicious domains but also to propose a practical solution to obtain large amounts of data amenable to various types of analysis. In order to position the numerous pieces of work that have followed, we propose a general framework (represented in Figure~\ref{fig:process}) to describe the various components required to implement a DNS based detection technique. It involves the following key components.

\smallbreak\noindent\textbf{DNS Data Collection.} DNS data could be collected at different locations of the DNS architecture as well as be available with different granularity. For example, they could be gathered at the recursive DNS server of a company or Internet Service Provider (ISP), or at higher-level authority servers. They may be available in the form of detailed DNS query/response logs, or only in aggregated forms. The location and granularity of the data can reveal different behaviors related to malicious domains, and thus have a significant impact on the intuition and design of the detection algorithms.
    
\smallbreak\noindent\textbf{Data Enrichment.} To get a more comprehensive view of malicious activities, DNS data often needs to be enriched by integrating networking and application data from various sources. Typical data sources used for this purpose include domain registration records, autonomous system numbers and geo-location information of IPs hosting domains.
    
\smallbreak\noindent\textbf{Algorithm Design.} A detection algorithm identifies a set of potentially malicious domains based on DNS data, enrichment information and, possibly, intelligence on existing known benign and malicious domains. Existing machine learning algorithms (supervised, semi-supervised and non-supervised) are often adapted in this context, relying on various intuitions about the behavior of malicious domains.
    
\smallbreak\noindent\textbf{Ground Truth.} A ground truth of malicious and benign domains is needed both in the algorithm design phase and in the evaluation phase. Supervised and semi-supervised detection algorithms rely on known malicious and benign domains to train a machine learning model and tune important parameters. The evaluation of detection algorithms is also greatly influenced by how the ground truth set is collected, cleaned and applied.
    
\smallbreak\noindent\textbf{Evaluation Methodology.} Malicious domain detection imposes unique challenges that are not observed in typical machine learning problems, including its highly dynamic nature and the adaptiveness of attackers. Therefore, besides following standard evaluation methodologies from the machine learning community, additional evaluation criteria and methods need to be adopted to reflect the true effectiveness of a malicious domain detection scheme in practice. For example, it is highly desirable to evaluate the robustness of the approaches against adaptive attackers who could change their behaviors deliberately to evade detection.
    
A DNS-based malicious domain detection technique can be characterized by the above five key components. Reading the relevant articles, we paid particular attention to the information about (i) the sources of DNS data, enrichment data, and ground truth data, (ii) the extracted features and how they are used in various approaches, (iii) the evaluation metrics and strategies. The collected information constitutes the core of Sections~\ref{sec:datasources},~\ref{sec:approaches}, and~\ref{sec:evaluation}. Reading the papers and using our domain expertise, we identified a number of problems and challenges in the area. In each of these sections, we also discuss the challenges or unsolved problems faced by the research community. In Section~\ref{sec:background} we provide necessary background on DNS and malicious activities leveraging DNS, while Section~\ref{sec:conclusion} concludes our discussion.

\section{Domain Name System Background}
\label{sec:background}
This section aims at setting a common background on DNS and providing necessary shortcuts. Readers who are familiar with the topic can probably skip this section and move on to Section~\ref{sec:datasources}.

\subsection{Domain Name System Operation}
\label{subsec:dns_operation}
The \textit{Domain Name System} is a hierarchical decentralized naming system that decouples the physical location (i.e., IP address) of a service and its logical address (i.e., its domain name), so that one can connect to the service using only its domain name. The DNS protocol has been introduced in November 1983 (IETF RFCs 882~\cite{RFC0882} and 883~\cite{RFC0883}, later superseded by RFCs 1034~\cite{RFC1034} and 1035~\cite{RFC1035} correspondingly), and now it is an essential part of the Internet infrastructure as exemplified by the attack in October 2016 against the DynDNS provider~\cite{DynDnsAttack}. By overloading their DNS servers with spurious requests, the attack prevented an extremely large portion of users from connecting to the Internet resources they needed access to. In this section, we briefly describe the key DNS concepts; the interested reader is referred to the IETF RFC 1034~\cite{RFC1034} for more details.

Domain names are organized as a suffix tree structure called domain namespace. The root of this tree is the domain called \emph{root} represented with a zero length label. The dot character is used in domain names to divide hierarchy levels. Parts between the dots are called labels. It should be noted that the trailing dot separating root domain is usually omitted; in the rest of the paper we follow this convention. The farthest right label is named the \emph{Top Level Domain} (TLD), such as ``com''. The domain directly on the left of a TLD is a \emph{Second Level Domain} (2LD), e.g., ``example.com''. \emph{Fully qualified domain name} (FQDN), e.g., ``www.example.com'', identifies a single node within the domain tree and is associated with the resource information composed of separate Resource Records (\emph{RR}s).  A Resource Record is defined by an \textbf{owner} (a domain name where the RR is found), a \textbf{type} field (an encoded 16 bit value that specifies the type of the RR), a \textbf{time-to-live} (TTL) value (time in seconds during which an RR should be cached), and an \textbf{RDATA} field, whose content and semantics depends on the value of the \textbf{type} field. In this work we are interested in the following types of RR (see~\cite{RFC1035} for the whole list): \texttt{A/AAAA} stores an IPv4/IPv6 host address; \texttt{NS} points to the authoritative server storing the information about the domain; \texttt{MX} is used to determine where a mail should be sent; \texttt{PTR} record maps a host's IP address to its domain name or host name.

The domain namespace information (in the form of \emph{resource records}) is stored in the hierarchical distributed database. Given the hierarchical structure, it is possible to divide it into separate \emph{zones} (all domains under a particular node) and delegate the control under them to different authorities, which maintain this information in \emph{zone files}. The Internet Corporation for Assigned Names and Numbers (ICANN)~\cite{ICANN}, a non-profit organization, is responsible for the creation of TLDs and delegation of their control to companies called \emph{registries}, who are in charge for all the domains ending with that particular TLD. Registries work in close collaboration with \emph{registrars}, companies like GoDaddy, which sell second level domains to domain owners (\emph{registrants}) and provide billing and customer support.

To query information from a DNS database, a client specifies in the request a domain name and what type of the resource record it wants to obtain. The main algorithm specifying how standard queries are processed, is described in the RFC 1034~\cite{RFC1034}. For the sake of this paper, it suffices to say (see also Figure~\ref{fig:process}) that clients in need of a domain name resolution, e.g., in need of knowing which IP corresponds to a given domain name, use the service of a \emph{resolver}. The resolver will run what is called a \emph{recursive} query on behalf of clients contacting it. This means that it will do its best to eventually return the needed response to the client and, to do so, may send a number of queries to various name servers without the client being involved. Once the resolver obtains an IP for a given domain name, it will cache the information, and, in most cases, will not query the same information for other clients. The important point to note here is that, due to caching, the resolver is the only one to have a complete view of how many clients use its services to resolve a given domain name. If a resolver does not have an answer in its cache for a given request, it will look for the \emph{authoritative name server} in charge of that domain name. This search usually starts by asking the so-called \emph{root name servers}. Authoritative name servers typically do not respond to recursive queries but, instead, to \emph{iterative} ones, providing information at their disposal and leave it up to the requester to continue his quest by following the lead provided.

\subsection{Domain Name System Security}
\label{subsec:dns_security}
DNS security has received a lot of attention from the research community over the years. There are plenty of attacks, where DNS is involved, and an even bigger number of methods to detect them. In this section, we briefly outline the area of \emph{DNS Security} splitting it into the four subareas entitled hereafter as follows: \emph{Securing DNS}, \emph{Securing Data Provided by DNS}, \emph{Securing Users from Attacks Leveraging DNS Disingenuously}, \emph{Securing Users from Attacks Leveraging DNS Genuinely}.

\smallbreak\noindent\textbf{Securing DNS}. Being a cornerstone technology of the Internet, all DNS components has been widely attacked and exploited by adversaries. The DNS infrastructure has been targeted by a number of denial of service attempts, the latest major case being the already mentioned attack on the DynDNS infrastructure~\cite{DynDnsAttack}. The DNS software has been the subject of attacks for many years now. According to Hoglund and McGraw~\cite{ExploitingSoftware_Hoglund2004}, one of the very first reported Linux worms, the ADM worm, was spreading in a stealthy way in 1999 thanks to a buffer overflow vulnerability in DNS servers. Better software security engineering techniques, large amount of replicas for key DNS servers, deployment of anti-DDoS mitigation tools are among the various solutions that have, quite successfully, been brought forward to secure DNS from these attacks.

\smallbreak\noindent\textbf{Securing Data Provided by DNS}. Attackers always try to subvert the data provided by legit DNS servers because this allows them to redirect traffic to controlled resources. In June 2008, two of the world's most important Internet regulatory web sites, ICANN and IANA, were hijacked~\cite{IcannAndIanaSitesHacked_Kravets2008}, which led to the creation of a set of best practices~\cite{SAC40_SSAC2009} that registrars should implement in order to keep the domain names of their customers secure. DNS hijacking and DNS poisoning attacks had been known for more than 25 years, with the seminal work by Steven Bellovin, produced in 1990 but withhold from publication until 1995~\cite{UsingDomainNameSystemForSystemBreakIns_Bellovin1995} followed by the 2002 birthday paradox attack~\cite{VUNote457875}. However, only in 2008 with the so-called Kaminsky's attack~\cite{VUNote800113}, people really started paying attention to them. A number of approaches have been proposed to detect such attacks but the ultimate protection comes with the ever wider deployment of DNSSEC. Attackers have also misused the popularity of some web sites by (re-)registering their domain names just after their expiration dates, usually taking advantage of the oversight of their legitimate owners who failed to renew the registration in due time~\cite{ThisGuyBoughtGoogle_Carson2016}. In this case, attackers exploit what is usually known as the ``residual trust'' of these stolen domains~\cite{DomainZ_Lever2016,  WhoisLostInTranslation_Lauinger2016}, collecting money from ads showed to regular customers of the domain, hijacking emails, or pushing malicious content to the fooled client machines~\cite{AllYourDnsRecordsPointToUs_Liu2016, WhoisLostInTranslation_Lauinger2016, DomainZ_Lever2016}. Another group of attacks falling into this category is generally known as cybersquatting~\cite{Cybersquatting_Wright2012}, when an attacker registers an Internet domain name somehow similar to a victim's domain name. Typosquatting is one of such attacks that exploits common mistakes made by the users when they type domain names in an address bar. Being very typical, this attack has received a lot of attention from the research community~\cite{SUT_Banerjee2011, TheLongTaile_Szurdi2014, SevenMonthsWorthOfMistakes_Agten2015, EverySecondsCounts_Khan2015}. Other attacks of this group include bitsquatting~\cite{Bitsquatting_Nikiforakis2013}, soundsquatting~\cite{Soundsquatting_Nikiforakis2014}, and combosquatting~\cite{HidingInPlainSight_Kintis2017}.

\smallbreak\noindent\textbf{Securing Users from Attacks Leveraging DNS Disingenuously}. A third class of DNS security threat has to do with the disingenuous use of the protocol. DNS is among the very few protocols allowed in, probably, every computing network. Not surprisingly, a number of malware samples and botnets have misused it to enable the communications between compromised hosts and their command and control servers. Various approaches have been exploited, e.g., by using some well known fields such as the free form TXT field, or by encoding commands in queried domain names~\cite{OnBotnetsThatUseDnsForCommandAndControl_Dietrich2011}. The same techniques have been also used for data exfiltration and malicious payload distribution~\cite{DetectionOfMaliciousPayloadDistributionChannelsInDns_MertKara2014}. Last but not least, in a more recent past, attackers have taken advantage of the fact that DNS replies are UDP based and much larger than the queries sent. They leverage these features to mount denial of service attacks, dubbed Reflective Denial of Service attacks~\cite{AmplificationHell_Rossow2014}, in which genuine DNS servers are being used to flood victims with large amounts of unwanted replies. 

\smallbreak\noindent\textbf{Securing Users from Attacks Leveraging DNS Genuinely}. In this survey we concentrate on the attacks that leverage DNS genuinely to make them more resilient using the properties and features of the DNS protocol. To run malicious campaigns, mischievers need various kinds of services hosted in remote servers. In the early days, it was a common practice for malware to hardcode the IP addresses of the servers to receive orders or to exfiltrate data. That practice was abandoned very rapidly because the capture of a single malware sample could lead to the extraction of all these IPs, what was enough to shut down the whole botnet. It became clear that these servers needed to be able to move across the IP space. This is exactly what DNS had been made for. 
Moreover, in order not to be blacklisted, domain names should also have to move across the domain name space. There are two main techniques that are used to achieve this agile behavior: \textbf{Domain-Flux} and \textbf{IP-Flux} (or \textbf{Fast-Flux}). The former refers to the strategy having several FQDNs associated with one IP address. Using a \textit{Domain Generation Algorithm} (DGA), a malware is able to dynamically generate new domain names (see~\cite{TaxonomyOfDga_Sood2016} for a taxonomy on DGAs), usually as a function of the date and time. This technique makes it difficult, short of having reverse engineered the DGA, to block the domain names used by a given botnet since these domains have a very short lifetime. The latter (IP-Flux) is characterized by the continuous change of IP addresses associated with a particular domain name. In this case, a malware builds a \textit{Fast Flux Service Network} (FFSN)~\cite{MeasuringAndDetectingFastFluxServiceNetworks_Holz2008} consisting of hundreds or even thousands of IP addresses assigned to a given domain name. When such a domain is queried, it is resolved to these IPs, which are frequently changed, thus, protecting the real location of the malicious service. Usually, the large pool of rotating IP addresses are not the final destination of the request for the content, they are just stopovers, so to speak, to reach the final destination, possibly after several other stops. \textit{Double-flux} networks are a more complex technique providing an additional layer of redundancy. Specifically, both the DNS A record sets and the authoritative NS records for a malicious domain are continually changed in a round robin manner and advertised into the fast flux service network. Clearly, these techniques can also be used in combination providing many-to-many relationship between FQDNs and IP addresses. 

Although these techniques are aligned with the specification of the DNS protocol, malware have abused them in various ways to improve the mobility of their servers and, thus, their resilience. The good news is that these techniques leave traces within DNS data. Such traces give researchers important clues to develop detection approaches taking into account the changes in domain-IP mappings using the unique viewpoint provided by the observation of the DNS traffic. In this survey, we focus on the approaches that are designed to detect domains involved in such malicious activities through the analysis of the relevant traces left in DNS data.
\section{Data Sources Definitions}
\label{sec:datasources}
In this section we categorize different types of DNS data, auxiliary information and ground truth that are used in the schemes proposed in the literature. The way these data are collected has a significant impact on the underlying assumptions and intuitions of malicious domain detection schemes. Table~\ref{tab:datasources} presents a short summary of this section and itemizes relevant articles. Notice that the table is not exhaustive, it only includes the most relevant examples of sources and articles.

\begin{table*}[t!]
\caption{Summary of the ``Data Sources Definitions'' Section}
\label{tab:datasources}

\footnotesize
\begin{minipage}{\linewidth}
\setcounter{mpfootnote}{\value{tab_footnote_cnt}}

\centering
\begin{tabular}{|p{10mm}|p{38mm}|p{76mm}|}
\hline
\multicolumn{1}{|c|}{\textbf{Component}} & \multicolumn{1}{c|}{\textbf{Dimension}} & \multicolumn{1}{c|}{\textbf{Sources and Related Works}} \\ \hline

DNS\newline Data
&
\textit{1. Where the Data is Collected}\newline
\tabindent a) Host-Resolver\newline
\tabindent b) DNS-DNS
&
\textbf{1a}: BDS~\cite{BotnetDetectionBasedOnDnsRecordsAndActiveProbing_Prieto2011}, Choi et al.~\cite{BotnetDetectionByMonitoringGroupActivitesInDnsTraffic_Choi2007}, BotGAD~\cite{BotGAD_Choi2009,BotGAD_Choi2012}, Lee and Lee~\cite{TrackingMultipleCCBotnetsByAnalyzingDnsTraffic_Lee2010}, Krishnan et al.~\cite{CrossingTheThreashold_Krishnan2013}, Manadhata et al.~\cite{DetectingMaliciousDomainsViaGraphInference_Manadhata2014}, Yadav et al.~\cite{DetectingAlgorithmicallyGeneratedMaliciousDomainNames_Yadav2010,DetectingAlgorithmicallyGeneratedDomainFluxAttacks_Yadav2012}, Smash~\cite{Smash_Zhang2015}, Segugio~\cite{Segugio_Rahbarinia2015,Segugio_Rahbarinia2016}, Perdisci et al.~\cite{DetectingMaliciousFluxServiceNetworks_Perdisci2009}, Oprea et al.~\cite{DetectionOfEarlyStageEnterpriseInfection_Oprea2015}, Stalmans and Irwin~\cite{FrameworkForDnsBasedDetectionAndMitigation_Stalmans2011} \newline
\textbf{1b}: Exposure~\cite{Exposure_Bilge2011,Exposure_Bilge2014}, Notos~\cite{Notos_Antonakakis2010}, Khalil et al.~\cite{GuiltyByAssociation_Khalil2016},  Kopis~\cite{Kopis_Antonakakis2011}, Huang and Greve~\cite{LargeScaleGraphMiningForWebReputationInference_Huang2015}, Yu et al.~\cite{SemiSupervisedTimeSeriesModeling_Yu2014}, Gao et al.~\cite{ReexaminingDnsFromAGlobalRecursiveResolverPerspective_Gao2016}, Mishsky et al.~\cite{TopologyBasedFlowModel_Mishsky2015}
\\ \cline{2-3}

&
\textit{2. How the Data is Collected}\newline
\tabindent a) Active\newline
\tabindent b) Passive
&
\textbf{2a} (\textit{Sources: Thales~\cite{Thales_Kountouras2016}}): Holz et al.~\cite{MeasuringAndDetectingFastFluxServiceNetworks_Holz2008}, Fluxor~\cite{Fluxor_Passerini2008}, Nazario and Holz~\cite{AsTheNetChurns_Nazario2008}, BDS~\cite{BotnetDetectionBasedOnDnsRecordsAndActiveProbing_Prieto2011}, Konte et al.~\cite{DynamicsOfOnlineScamHostingInfrastructure_Konte2009}, Ma et al.~\cite{BeyondBlacklists_Ma2009}, Felegyhazi et al.~\cite{OnThePotentialOfProactiveDomainBlacklisting_Felegyhazi2010}, Hao et al.~\cite{MonitoringInitialDnsBehaviorOfMaliciousDomains_Hao2011}, Predator~\cite{Predator_Hao2016}, DomainProfiler~\cite{DomainProfiler_Chiba2016} \newline
\textbf{2b} (\textit{Sources: Farsight database~\cite{DNSDB}}): Choi et al.~\cite{BotnetDetectionByMonitoringGroupActivitesInDnsTraffic_Choi2007}, BotGAD~\cite{BotGAD_Choi2009,BotGAD_Choi2012}, Manadhata et al.~\cite{DetectingMaliciousDomainsViaGraphInference_Manadhata2014}, Exposure~\cite{Exposure_Bilge2011,Exposure_Bilge2014}, Notos~\cite{Notos_Antonakakis2010}, Khalil et al.~\cite{GuiltyByAssociation_Khalil2016},  Kopis~\cite{Kopis_Antonakakis2011}, Huang and Greve~\cite{LargeScaleGraphMiningForWebReputationInference_Huang2015}, Yu et al.~\cite{SemiSupervisedTimeSeriesModeling_Yu2014}, Gao et al.~\cite{ReexaminingDnsFromAGlobalRecursiveResolverPerspective_Gao2016},  Yadav et al.~\cite{DetectingAlgorithmicallyGeneratedMaliciousDomainNames_Yadav2010,DetectingAlgorithmicallyGeneratedDomainFluxAttacks_Yadav2012}, Smash~\cite{Smash_Zhang2015}, Segugio~\cite{Segugio_Rahbarinia2015,Segugio_Rahbarinia2016}, FluxBuster~\cite{FluxBuster_Perdisci2012}, Perdisci et al.~\cite{DetectingMaliciousFluxServiceNetworks_Perdisci2009}, Oprea et al.~\cite{DetectionOfEarlyStageEnterpriseInfection_Oprea2015}, Mishsky et al.~\cite{TopologyBasedFlowModel_Mishsky2015}
\\ \hline

Data\newline Enrichment
&
\textit{1. Type of the Enrichment Data}\newline
\tabindent a) Geo-location\newline
\tabindent b) ASN\newline
\tabindent c) Registration records\newline
\tabindent d) IP/domain black-/whitelists\newline
\tabindent e) Associated resource records\newline
\tabindent f) Network data
&
\textbf{1a} (\textit{Sources: MaxmindDB~\cite{MaxmindDbs}}): Seifert et al.~\cite{IdentificationOfMaliciousWebPages_Seifert2008}, Exposure~\cite{Exposure_Bilge2011, Exposure_Bilge2014}, BotGAD~\cite{BotGAD_Choi2009,BotGAD_Choi2012}, Gao et al.~\cite{EmpiricalReexaminationOfGlobalDnsBehavior_Gao2013, ReexaminingDnsFromAGlobalRecursiveResolverPerspective_Gao2016}, Zou et al.~\cite{DetectingMalwareBasedOnDnsGraphMining_Zou2015} \newline
\textbf{1b} (\textit{Sources: MaxmindDB~\cite{MaxmindDbs}, Team Cymru~\cite{TeamCymru}}): Khalil et al.~\cite{GuiltyByAssociation_Khalil2016}, DomainProfiler~\cite{DomainProfiler_Chiba2016}, Fukuda and Heidemann~\cite{DetectingMaliciousActivityWithDnsBackscatter_Fukuda2015}, Stevanovic et al.~\cite{OnTheGroundTruthProblem_Stevanovic2015}, Kopis~\cite{Kopis_Antonakakis2011} \newline
\textbf{1c} (\textit{Sources: WHOIS~\cite{RFC3912}, commercial~\cite{Who_is, DomainHistory, DomainTools}}): Felegyhazi et al.~\cite{OnThePotentialOfProactiveDomainBlacklisting_Felegyhazi2010}, Predator~\cite{Predator_Hao2016}, Fluxor~\cite{Fluxor_Passerini2008} \newline
\textbf{1d} (\textit{Sources: see ground truth}): Notos~\cite{Notos_Antonakakis2010}, Prieto et al.~\cite{BotnetDetectionBasedOnDnsRecordsAndActiveProbing_Prieto2011} \newline
\textbf{1e} (\textit{Sources: Thales~\cite{Thales_Kountouras2016}, Farsight DNS database~\cite{DNSDB}}): Hao et al.~\cite{MonitoringInitialDnsBehaviorOfMaliciousDomains_Hao2011}, Prieto et al.~\cite{BotnetDetectionBasedOnDnsRecordsAndActiveProbing_Prieto2011} \newline
\textbf{1f} (\textit{Sources: Censys~\cite{Censys_Durumeric2015}, Shodan~\cite{Shodan}, Team Cymru~\cite{TeamCymru}}): Nadji et al.~\cite{ConnectedColors_Nadji2013}, Prieto et al.~\cite{BotnetDetectionBasedOnDnsRecordsAndActiveProbing_Prieto2011}
\\ \hline

Ground\newline Truth
&
\textit{1. Type of the Ground Truth}\newline
\tabindent a) Malicious\newline
\tabindent b) Benign
&
\textbf{1a}: Spamhaus~\cite{Spamhaus}, Yahoo Webspam Database~\cite{YahooWebspamDatabase}, PhishTank~\cite{PhishTank}, VirusTotal~\cite{VirusTotal}, McAfee SiteAdvisor~\cite{McafeeSiteAdvisor}, Malware Domains~\cite{MalwareDomains}, Malware Domains List~\cite{MalwareDomainList}, UrlVoid~\cite{UrlVoid}, Wepawet~\cite{Wepawet}, McAfee SiteAdvisor~\cite{McafeeSiteAdvisor}, Google Safe Browsing~\cite{GoogleSafeBrowsing}, Web Of Trust~\cite{WOT}, Anubis~\cite{Anubis} \newline
\textbf{1b}: Alexa top ranked domains~\cite{Alexa}, McAfee SiteAdvisor~\cite{McafeeSiteAdvisor}, Google Safe Browsing~\cite{GoogleSafeBrowsing}, Web Of Trust~\cite{WOT}
\\ \hline

\end{tabular}
\setcounter{tab_footnote_cnt}{\value{mpfootnote}}
\end{minipage}
\end{table*}

\subsection{Sources of DNS Data}
\label{subsec:dns_data}

The collection of DNS data can be categorized along the following two orthogonal dimensions: (1) \emph{where} and (2) \emph{how} the data is collected.

\smallbreak\noindent\textbf{Where the Data is Collected}. Due to the distributed nature of the DNS infrastructure, multiple locations can be considered to collect information about DNS queries and replies. Among all servers involved, the resolver (as defined in Section~\ref{sec:background}) is unique as it is the only location which has access to queries coming directly from client machines. Therefore, in the following, we distinguish two specific cases for the sources of the data. We call the first one ``\textbf{Host-Resolver}''. It refers to DNS data obtained by observing the communications between an end host and its resolver. The second is called ``\textbf{DNS-DNS}'' and refers to the data that can be obtained by observing the communications between two DNS servers (and one of them could, possibly, be a resolver).

\smallbreak\noindent\textbf{How the Data is Collected}. Obtaining information about existing associations between IPs and domain names at a given point of time can be done in two ways. One way is to resolve \textbf{actively} and regularly a large collection of domain names to obtain that information. Another way is to observe \textbf{passively} all the requests sent to DNS servers extracting the necessary data. In the following, we will distinguish these two methods as \textbf{Active vs. Passive DNS data collection.}

\subsubsection{Where the Data is Collected}
\paragraph{Host-Resolver (Flows 1 and 8 in Figure~\ref{fig:process})}
One major advantage of the data captured at the internal interface of a resolver is that it provides detailed information about the clients in terms of DNS queries and responses, which may directly link to certain types of malicious behaviors~\cite{BotnetDetectionBasedOnDnsRecordsAndActiveProbing_Prieto2011, BotnetDetectionByMonitoringGroupActivitesInDnsTraffic_Choi2007, BotGAD_Choi2012, DetectingMaliciousDomainsViaGraphInference_Manadhata2014, Segugio_Rahbarinia2015, Segugio_Rahbarinia2016, DetectionOfEarlyStageEnterpriseInfection_Oprea2015}. For example, hosts controlled by a botnet often have similar DNS query model in terms of both queried domains and temporal patterns. Choi et al.~\cite{BotGAD_Choi2009, BotGAD_Choi2012} use the information ``what host queries what domain'' to build a matrix for every domain that shows what machine at what period of time has queried this particular domain. Such a representation is very handy because it allows the analysts to prune matrix both column-wise and row-wise to correct errors that could arise due to deactivation of a botnet part or if time-window parameter is misconfigured. In the Segugio system~\cite{Segugio_Rahbarinia2015, Segugio_Rahbarinia2016}, this information is used to build a host-domain graph representing ``who-queries-what'' relation between hosts and domains. It would be harder to observe such behavior patterns from DNS-DNS data due to caching by the intermediate servers. Another advantage of this source of data is the ease of access. Any company or research institute could directly deploy sensors at its own resolver(s) requiring no co-operation with other parties. Due to these reasons, many existing schemes for malicious domain detection are built on data from resolvers, in particular those whose features are tied to the behavior of individual hosts. It should be also mentioned that the approaches that use Host-Resolver DNS data, may be also adapted to detect malicious hosts, roughly speaking those ones which query malicious domains.

One limitation of sensors deployed at the internal interface of a resolver is that they can only see the behavior of hosts inside a single organization, which may not be comprehensive enough to establish patterns related to malicious activities. One notable exception is when the client chooses to use, as a resolver, a publicly available DNS server willing to serve recursive queries, such as Google Public DNS~\cite{GooglePublicDns}, OpenDNS~\cite{OpenDNS}, or Norton ConnectSafe~\cite{NortonConnectSafe}. Due to the sheer volume and diversity of hosts they interact with, the data collected at these resolvers is suitable to comprehensively reveal suspicious behaviors related to different kinds of attacks. The DNS resolvers of large ISPs also serve a large amount of individual users. They can be used for the same purpose. Unfortunately, DNS data logs from public DNS servers or ISP DNS servers are not easily accessible to the research community, often because of privacy concerns~\cite{AnalysisOfPrivacyDisclosure_Zhao2008, BehaviorBasedTracking_Herrmann2013, TrackedWithoutTrace_Kirchler2016}.

\paragraph{DNS-DNS (Flows 2 to 7 in Figure~\ref{fig:process})}
On the other hand, queries observed by sensors deployed near other DNS servers usually see queries issued from several organizations. In the literature, the most frequent locations considered to observe DNS-DNS traffic are (i) at the authoritative name servers~\cite{Kopis_Antonakakis2011} including the servers responsible for TLDs~\cite{KinderedDomains_Thomas2014, Kopis_Antonakakis2011}, and (ii) at the external interface of the resolvers~\cite{Exposure_Bilge2011,Exposure_Bilge2014,Notos_Antonakakis2010,GuiltyByAssociation_Khalil2016}. The closer the sensor to the roots of the DNS tree, the larger the visibility. The data collected from TLD servers could offer unique insights and early detection of newly emerged malicious domains. 
Note, that such logs would only reveal the existence of the requests but not the answers to them (i.e., the IPs requested) since TLD servers typically serve only iterative queries. Such signals would be hard to capture only from logs of resolvers. Getting logs from an authoritative server solves this issue but, due to caching, not all queries will be visible to that server. Therefore, the view provided by the logs of DNS servers higher in the DNS tree can quickly become rather coarse grained. The extreme case is the requests observed at the root servers which give almost a full visibility of all names queried over the Internet but none of the responses. Volumetric analysis of these requests is also heavily impacted by caching happening in the intermediate servers between the end clients and the root servers~\cite{EmpiricalReexaminationOfGlobalDnsBehavior_Gao2013,ReexaminingDnsFromAGlobalRecursiveResolverPerspective_Gao2016}. Therefore, the features offered by the data captured at servers different from the resolvers are often limited. Furthermore, the logs from such domain servers cannot be easily obtained by researchers.

\subsubsection{How the Data is Collected}
\paragraph{Active DNS data collection}
To actively obtain DNS data, a data collector would deliberately send DNS queries and record the corresponding DNS responses~\cite{AsTheNetChurns_Nazario2008, MeasuringAndDetectingFastFluxServiceNetworks_Holz2008, Fluxor_Passerini2008, DynamicsOfOnlineScamHostingInfrastructure_Konte2009, BeyondBlacklists_Ma2009, DomainProfiler_Chiba2016, Thales_Kountouras2016}. The list of queried domains is built thanks to multiple sources, typical ones include popular domains lists such as the Alexa Top Sites~\cite{Alexa}, domains appearing in various blacklists, or those from the zone files of authoritative servers. Clearly, as the queries are issued by the data collector, they do not reflect the behavior of actual users. Instead, active DNS data mainly capture the DNS records of domains, e.g., the resolved IPs, canonical names, TTL of a record, etc. The major advantages of actively crawled DNS data are the flexibility and ease of use of the data collection method. Data collectors can easily control which domains to query. Additionally, active DNS can reveal abuse signals about domains before their actual malicious use. For example, active DNS collector can discover in zone files a potentially malicious domain that has been newly registered but not yet used~\cite{OnThePotentialOfProactiveDomainBlacklisting_Felegyhazi2010, Predator_Hao2016}, while passive sensors cannot see it yet. Moreover, active DNS data are not linked to the behavior of individual users, and therefore, can be shared with the research community without any privacy concern.  Meanwhile, due to the same reason, active DNS data could not be used to detect malicious domains with techniques that rely on user-level features (e.g., temporal statistics of user queries). If the DNS queries are issued only from a limited set of hosts, the collected data could be biased, and this is another limitation. Specifically, a domain could be associated to multiple IPs depending on the geo-location of the query issuer. Therefore, active DNS data may contain a limited small set of IPs that are a function of where the queries are issued.

\paragraph{Passive DNS data collection}
Collecting DNS data passively is done by deploying sensors in front of DNS servers or by having access to DNS server logs to obtain real DNS queries and responses~\cite{BotGAD_Choi2012, DetectingMaliciousDomainsViaGraphInference_Manadhata2014, Exposure_Bilge2011, Exposure_Bilge2014, Notos_Antonakakis2010, GuiltyByAssociation_Khalil2016, Kopis_Antonakakis2011, Segugio_Rahbarinia2015, Segugio_Rahbarinia2016, DetectionOfEarlyStageEnterpriseInfection_Oprea2015}. Therefore, DNS data collected passively are more representative and more ``revealing'' in sense of a rich set of features and statistics that could be derived to identify malicious activities. For instance, the Kopis system~\cite{Kopis_Antonakakis2011} in order to build a requester profile and to assess requester diversity, requires information regarding every resolver that queried data about a particular domain from an authoritative or TLD DNS server. Further, if sensors are deployed in DNS servers of diverse organizations from different locations, DNS data collected passively are likely to be more comprehensive than the ones collected actively. This assumption is indirectly confirmed by Rahbarinia et al.~\cite{Segugio_Rahbarinia2015, Segugio_Rahbarinia2016}. Their system performed better if training and testing was executed on the data from the same ISP than if obtained from different ISPs. Moreover, such approaches do not require an initial precompiled list of domains. On the other hand, sharing of such data could be hindered due to privacy concerns, especially if sensors are deployed between clients and resolvers. Therefore, the existing publicly available passive DNS datasets are collected after resolvers and usually provide only aggregated views of queries to hide individual activities. For example, the Farsight passive DNS database~\cite{DNSDB} does not contain the IP addresses of requesters. Furthermore, for a given domain and one of its resolved IPs, it offers only the timestamps of the first and last seen resolution and the total number of them in between. This is a trade-off between privacy protection, the ability of sharing and the utility of the data. Similar to the actively obtained DNS data, it would be impossible to build fine-grained user-level features from this dataset. However, we note that, as the aggregation is done over queries and responses due to the actual host/user activities, some important aggregated user statistics still could be derived that may be very useful for malicious domain detection. For example, it is still possible to observe a sudden increase of queries over a set of domains globally in a short period of time, even after aggregation. Such statistics would not be available from actively collected DNS data.

\subsubsection{Challenges}~\\
The challenges of access to DNS data faced by the research community lie in two aspects. The first is in the data collection phase. Though DNS traffic is present in all networks, collection of a datasets is not an easy task. As discussed earlier, it is relatively easy to set up DNS traffic sensors in a single organization's network (e.g., a campus network), but then the collected data could offer only a limited local view of global threats. The peculiarity of many existing DNS-based malicious domain detection techniques is that they work best in big data scenarios. Thus, they may not be able to produce meaningful results on datasets collected in small networks. Meanwhile, integrating data from DNS servers belonging to different organizations would often face significant bureaucratic/legal obstacles, due to the sensitive nature of DNS logs. The same is true if researchers would like to gain access to the data from public DNS servers or from ISPs.

Even a bigger challenge lies in data sharing. Unfortunately, security related data are notoriously sensitive and hard to share. Even if a researcher is able to gain access to DNS logs from an ISP, it would be extremely difficult to make the same data available to peers for validation. At the same time, scientific advances rely on validation of and comparison with the existing approaches. There were some attempts to compare new approaches with the previous ones (e.g., Rahbarinia et al.~\cite{Segugio_Rahbarinia2015} compared their approach with Notos~\cite{Notos_Antonakakis2010}), but current research significantly lacks extensive and systematic experimental validation and comparison of different techniques. The primary reason lies in the difficulty to make publicly available a set of common or comparable reference datasets. Although currently there are several publicly available DNS datasets, which have been collected passively (e.g., from Farsight~\cite{DNSDB}) or actively (e.g., Thales~\cite{Thales_Kountouras2016}), they cannot be used in many approaches, especially in those relying on client-side patterns~\cite{DetectionOfEarlyStageEnterpriseInfection_Oprea2015, DetectingMaliciousDomainsViaGraphInference_Manadhata2014}. It should be also noted that despite some approaches may work on data collected both actively and passively (for instance, the one proposed by Khalil et al.~\cite{GuiltyByAssociation_Khalil2016} which relies on domain co-location information obtainable from both datasets), such a comparison has never been performed before.

Moreover, the researchers must ensure that the results obtained with a particular dataset can be generalized to all other possible datasets. Clearly, some datasets may have space or time peculiarities that can influence the results considerably. For instance, Yadav et al.~\cite{DetectingAlgorithmicallyGeneratedMaliciousDomainNames_Yadav2010, DetectingAlgorithmicallyGeneratedDomainFluxAttacks_Yadav2012} grounded their approach on the insight that the domain names generated automatically have abnormal distribution of character frequencies and that the algorithmically produced names are usually unpronounceable for an english speaker. Although in general this may be true for the majority of domain names, there are many countries in the world, e.g., China or Russia, where such intuitions may not hold true. It may be required to adjust the model to the peculiarities of the region.

\subsection{Sources of Data Enrichment}
\label{subsec:data_enrichment}
DNS data represents an important source of intelligence that has been successfully used by many approaches to discover and predict malicious activities. However, to provide deeper insights about malicious activities and to enhance the accuracy and coverage, the majority of the detection approaches presented in this survey utilizes external sources of data to enrich DNS information. For example, mapping the IP address to a hosting country enables some approaches to use the trustworthiness of the country as a feature in classifying the maliciousness of domains/IPs~\cite{OnTheGroundTruthProblem_Stevanovic2015}. Generally, the sources of data enrichment can be classified by the \textbf{Type of Information} they provide.

\subsubsection{Enrichment Information Types}
\paragraph{Geo-location} The geo-locations of IPs and domains are commonly used to understand the diversity of the origins of the DNS queries as well as of machines hosting the domains. Such kind of enrichment is seen in a large number of papers, e.g., in~\cite{Exposure_Bilge2011, Exposure_Bilge2014, Fluxor_Passerini2008, OnTheGroundTruthProblem_Stevanovic2015, TopologyBasedFlowModel_Mishsky2015}. The most common source of IP geolocation information observed in the literature is the Maxmind database~\cite{MaxmindDbs}.

\paragraph{Autonomous system number (ASN)} This source of information enables to understand the distribution and utilization of adversary resources~\cite{MonitoringFastFluxBotnetUsingRecursiveAndPassiveDns_Mahjoub2013, Exposure_Bilge2014, TopologyBasedFlowModel_Mishsky2015}. For example, legitimate domains (except those using CDNs) are usually hosted on one or few ASNs as opposed to malicious domains which hop from one ASN to another to evade detection. ASN is a valuable source of information allowing to distinguish different types of Internet services (e.g., IPs only used by dedicated organizations vs. those belonging to cloud service providers). The information on the IP-ASN mapping can be found in the Maxmind database~\cite{MaxmindDbs} or using the Team Cymru service~\cite{TeamCymru}.

\paragraph{Registration records} Even though domain registration records often are not verified by authorities, the information located there sometimes can be used as supportive evidence to link malicious domains controlled by the same adversary. Further, temporal information of registration records (e.g., their creation/expiration time) is critical to identify domains registered automatically in bulks to be used later for malicious activities. In fact, some previous works rely purely on registration records to identify malicious domains~\cite{OnThePotentialOfProactiveDomainBlacklisting_Felegyhazi2010, UnderstandingTheDomainRegistrationBehaviorOfSpammers_Hao2013, Predator_Hao2016}. The registration records information is usually obtained from servers which provide access to it through the WHOIS protocol~\cite{RFC3912}. It should be mentioned, there is no common standard on the format of the data provided. Hence, researchers must develop custom parsers in order to extract the necessary data.

\paragraph{IP/domain blacklists/whitelists} Domains are also often checked against well-known IP/domain blacklists (more information about blacklists/whitelists will be given in Section~\ref{subsec:sources_ground_truth}). For example, Notos~\cite{Notos_Antonakakis2010} checks how many of the IPs associated to a domain are blacklisted, which is expected to be an indicator of the maliciousness of this domain. Other approaches check if the related IPs/domains are blacklisted. For instance, Prieto et al.~\cite{BotnetDetectionBasedOnDnsRecordsAndActiveProbing_Prieto2011} considers a domain suspicious if its authoritative name server is blacklisted.

\paragraph{Associated resource records} It is possible to gain more information about a given domain or IP by exploring other \textsc{RRs} related to it that can be retrieved from the DNS database. For instance, Hao et al.~\cite{MonitoringInitialDnsBehaviorOfMaliciousDomains_Hao2011} have shown that the distribution of DNS MX records in the IP space for malicious domains is different than that of the benign ones. Moreover, Prieto et al.~\cite{BotnetDetectionBasedOnDnsRecordsAndActiveProbing_Prieto2011} observed that domain names associated with a botnet usually do not have any associated MX record. 

\paragraph{Network data} The IP/domain data can be also enriched with information from network activities~\cite{BotnetDetectionBasedOnDnsRecordsAndActiveProbing_Prieto2011}, e.g., if a website is associated with a domain, what is the HTTP response, what ports are opened, etc. Researchers usually obtain such kind of information by developing their own probes or using the information provided by Internet-wide scanners such as Censys~\cite{Censys_Durumeric2015} or Shodan~\cite{Shodan}.

\subsubsection{Challenges}~\\
It is important to understand that the information associated with an IP or a domain does vary over time. For instance, the Maxmind database~\cite{MaxmindDbs}, which is used to enrich data with the geolocation and ASN information, is frequently updated. Therefore, the values of the features calculated using these data also change. This results in a number of challenges. First, since researchers often work with historical DNS data, they must rely on the enrichment information available at the same time frame when the DNS data was collected. For instance, if they calculate the number of countries that host a particular domain at a given date, they have to use the information from the Maxmind database available at exactly the same date. As an alternative, they can use the most recent available enrichment data. Any of the approaches may be valid, and researchers must clearly identify which is used. The second challenge is tightly connected to the first one. Given a large number of IP addresses, the fast-growing number of domain names and the frequent change of the corresponding enrichment data, the maintenance and management of the related historical information requires a lot of resources that may not be available for researchers.

\subsection{Sources of Ground Truth}
\label{subsec:sources_ground_truth}
Practically, every approach to detect malicious domains requires high-quality ground truth for training and validation. The ground truth data in this area is associated with domains and can be divided according to the \textbf{Type}.

\subsubsection{Type of Ground Truth}
\paragraph{Malicious Ground Truth}
To get a ground truth of malicious domains, the dominant practice in existing works is to extract it from various public blacklists. Some of the blacklists are only about specific malicious activities, e.g., spams (Spamhaus~\cite{Spamhaus}, Yahoo Webspam Database~\cite{YahooWebspamDatabase}), phishing (PhishTank~\cite{PhishTank}, OpenPhish~\cite{OpenPhish}), while some others are more general and include domains/IPs involved in any kind of malicious activities, e.g., VirusTotal~\cite{VirusTotal}, McAfee SiteAdvisor~\cite{McafeeSiteAdvisor}, Malware Domains~\cite{MalwareDomains} and Malware Domains List~\cite{MalwareDomainList}. Some of these sources, such as WoT~\cite{WOT}, can also blacklist domains that are not, per se, associated with malicious activities. This is the case when the content of such web sites is considered inappropriate with respect to the policies in place for the specific blacklist considered (e.g., pornographic content, violence, racism, copyrighted material, etc.). Another source to build ground truth is proprietary blacklist/whitelists, or proprietary reputation systems deployed by anti-virus security companies (e.g., Symantec), whose availability to the general research community is quite limited.

\paragraph{Benign Ground Truth}
Ground truth of benign domains in the literature is largely drawn from highly ranked popular domains. For example, Alexa top ranked domains~\cite{Alexa} are commonly used\footnote{We will explain later the need to apply a supplementary filter to the Alexa lists because they do contain malicious domains as well.}. Another common practice, at least when building an initial candidate set of benign domains, is based on the top level domains. For example, domains from ``gov'' and ``mil'' zones or those belonging to Google and Microsoft (used, e.g., in~\cite{DetectingMalwareBasedOnDnsGraphMining_Zou2015}), are generally considered more trustworthy than those from ``com'' or ``info''. Additionally, some public cyber intelligence tools like McAfee SiteAdvisor~\cite{McafeeSiteAdvisor}, Google Safe Browsing~\cite{GoogleSafeBrowsing} or Web Of Trust~\cite{WOT} report not only malicious and suspicious domains, but also benign ones and hence, can be also used to extract benign ground truth.

\subsubsection{Challenges}
\paragraph{Malicious Ground Truth Challenges} Even though reputable blacklists generally provide robust evidences about blacklisted domains, they still have a number of subtle  issues. First, a malicious domain can be malignant in different ways: spam, phishing, C\&C, unethical, adult content, etc. Thus, the mere definition of the term ``malicious'' differs from one ground truth dataset to another. The ground truth collected for one approach may not work for another one that focuses on detecting domains involved in other types of malicious activities. Second, blacklists employ different collection methods. For instance, they may rely on crowd sourced data (e.g., PhishTank~\cite{PhishTank}, Web of Trust~\cite{WOT}), may crawl and analyze website content (e.g., Wepawet~\cite{Wepawet}), may run malicious software in sandboxes and analyze accessed domains (e.g.,  Anubis~\cite{Anubis}), may reverse botnet protocol and generate feed of names produced by DGAs (e.g, Conficker~\cite{ContainingConficker_Honeynet2009}), may be obtained using internal tools (e.g., Google Safe Browsing~\cite{GoogleSafeBrowsing}) or may aggregate data from different sources (for instance, UrlVoid~\cite{UrlVoid} or VirusTotal~\cite{VirusTotal}). Third, none of the blacklists is completely reliable. Sinha et al.~\cite{ShadesOfGrey_Sinha2008} and Ramachandran et al.~\cite{CanDnsBasedBlacklistsKeepUpWithBots_Ramachandran2006} showed that blacklists exhibit high false positives and false negatives rates. Some approaches address this by cross-checking domains in multiple blacklists. For example, Kheir et al.~\cite{Mentor_Kheir2014} built a ground truth dataset by voting on 3 different blacklists.

\paragraph{Benign Ground Truth Challenges} Although blacklists may contain false positives, generally a domain can be considered as malicious if it has appeared in a reputable blacklist. At the same time, building a bening domain ground truth is a far harder task. A domain cannot be deemed as benign simply because it is not present in any known blacklist. The large number of Internet domains (according to Verisign~\cite{InternetGrows_Verisign2016}, in 2016 there were around 314 million 2LDs) makes it impossible to scan and check them regularly. Although this number is large, it represents only a very small portion of the total number of FQDNs in the Internet. Even worse, that number keeps growing every day. Therefore, a malicious domain may not be blacklisted because it did not expose malicious content when it was scanned or it has never been scanned.

Although the usage of top $K$ Alexa domains~\cite{Alexa} as benign ground truth makes sense (the administrators of popular web pages devote more effort to protect their resources), it is both limited and suffers from high false positive rate. The list contains only 2LD domains and does not provide any information about sub-domains, which makes it rather limited. Domains are ranked according to their popularity but not based on their security or safety, which leads to high false positive rate. It contains proxies to malicious web pages or even domains hosting malicious content. For instance, a quite popular 2LD, \url{unblocksit.es} (ranked 11550 as of April 1, 2016), offers to proxy access to other, potentially blacklisted, domains. This 2LD is not, per se, malicious since it can be used by legit users to try to circumvent censorship measures that they are facing. Similarly, malicious users can abuse this service as a safe haven to defeat known blocking mechanisms. Moreover, some malicious domains could appear among top $K$ Alexa domains due to a burst of requests from a high number of infected clients querying them. Stevanovic et al.~\cite{OnTheGroundTruthProblem_Stevanovic2015} cross-checked domains from Alexa top $K$ domains with UrlVoid~\cite{UrlVoid}, the service which aggregates information from different blacklists. The results show that a relatively high percentage of domains (around 15\% out of 10,000 top domains) is reported to be malicious by at least one blacklist.

Such impurities of benign ground truth negatively affect the accuracy of domain detection approaches. For instance, consider a malicious domain $d$ that is mislabeled as benign in the ground truth, as it is in Alexa top $K$ domains. A correct detection of $d$ would be counted as a false positive incorrectly, causing the measured false positive rate higher than what it really is. At the same time, a malicious domain with a strong association with $d$ may be missed due to the lack of associations with malicious domains, that negatively affects the true positive rate. To mitigate the impact of Alexa top $K$ impurities, some approaches filter the domains before adding them to benign ground truth. For example, Rahbarinia et al.~\cite{Segugio_Rahbarinia2015} consider only domains that consistently appear in Alexa top 1 million sites for one year. Similarly, Bilge et al.~\cite{Exposure_Bilge2011,Exposure_Bilge2014} consider only domains older than 1 year as benign. Some other approaches, e.g.~\cite{Notos_Antonakakis2010,GuiltyByAssociation_Khalil2016}, remove dynamic DNS service domains, such as \url{no-ip.com}, before building a ground truth of benign domains. As one can see, there is no consensus on what could or should constitute the ground truth for benign domains.

\paragraph{Common Challenges} One of the common issues is to understand what domain level to use for ground truth compilation: 2LD, 3LD, or FQDN. Some ground truth sources contain domains of a specific level, e.g., top $K$ Alexa domains~\cite{Alexa} mostly consists of 2LD domains. This creates trouble for approaches that focus on the domain levels different from those found in the ground truth. The relations between domains at different levels are also unclear. Should we consider any subdomain of a malicious/benign domain as malicious/benign? Should we consider a domain as malicious/benign if the majority of its subdomains are malicious/benign? Unfortunately, there is no definite ``Yes'' or ``No'' answer to these questions. It may be reasonable, to a certain extent, to answer ``Yes'' to these questions for 2LDs that belong to private organizations like Google or Facebook. However, the subdomains of dynamic DNS services such as \url{no-ip.com} and \url{3322.org} may be totally unrelated and hence, cannot be assumed benign even if the vast majority of their subdomains is benign.

Another common issue, which we have identified in the literature, is the limited quantitative discussion of training and testing sets comprising the ground truth data. It has been shown (see~\cite{DataMiningForImbalancedDatasets_Chawla2005,TheRoleOfBalancedTrainingAndTestingDatasets_Wei2013}) that an imbalanced training dataset may have considerable influence on the learning of a classifier and thus, may influence some of the measured metrics.

\section{Design of Detection Algorithms}
\label{sec:approaches}
To systematically present the approaches that have been used by the various lines of work in the past, we have opted to look at them from three distinct viewpoints:

\begin{description}
  \item [Features:] What features are used?
  \item [Method:] What technology is the detection method based on?
  \item [Outcome:] What outcome is produced?
\end{description}

The following subsections address each of these viewpoints separately, while Table~\ref{tab:approaches} provides an overview of the section by itemizing only the most relevant examples and related articles.

\begin{table*}[t!]
\caption{Summary of the ``Design of Detection Algorithms'' Section}
\label{tab:approaches}

\footnotesize
\begin{minipage}{\linewidth}
\setcounter{mpfootnote}{\value{tab_footnote_cnt}}

\centering
\begin{tabular}{|p{10mm}|p{38mm}|p{76mm}|}
\hline
\multicolumn{1}{|c|}{\textbf{Component}} & \multicolumn{1}{c|}{\textbf{Dimension}} & \multicolumn{1}{c|}{\textbf{Examples and Related Works}} \\ \hline

Features
&
\textit{1. Internal vs. Contextual}\newline
\tabindent a) Internal \newline
\tabindent b) Contextual
&
\textbf{1a} (\textit{Examples: Domain average TTL value, Domain name label features, Association-based features}): Exposure~\cite{Exposure_Bilge2011,Exposure_Bilge2014}, Perdisci et al.~\cite{DetectingMaliciousFluxServiceNetworks_Perdisci2009}, FluxBuster~\cite{FluxBuster_Perdisci2012}, Stalmans and Irwin~\cite{FrameworkForDnsBasedDetectionAndMitigation_Stalmans2011}, Notos~\cite{Notos_Antonakakis2010}, Pleiades~\cite{Pleiades_Antonakakis2012}, BotGAD~\cite{BotGAD_Choi2009,BotGAD_Choi2012}, Phoenix~\cite{Phoenix_Schiavoni2014}, Zou et al.~\cite{DetectingMalwareBasedOnDnsGraphMining_Zou2015}, Oprea et al.~\cite{DetectionOfEarlyStageEnterpriseInfection_Oprea2015}, GMAD~\cite{GMAD_Lee2014}, Segugio~\cite{Segugio_Rahbarinia2015,Segugio_Rahbarinia2016}, Stevanovic et al.~\cite{MethodForIdentifyingCompromisedClients_Stevanovic2017} \newline
\textbf{1b} (\textit{Examples: Number of ASNs to which the IP addresses of a domain belong to, Number of ASNs, Historical association of domains to IP addresses}): Chiba et al.~\cite{DomainProfiler_Chiba2016}, Stalmans and Irwin~\cite{FrameworkForDnsBasedDetectionAndMitigation_Stalmans2011}, BotGAD~\cite{BotGAD_Choi2009,BotGAD_Choi2012}, Hu et al.~\cite{MeasurementAndAnalysisOfGlobalIpUsagePatternsOfFastFluxBotnets_Hu2011}, Khalil et al.~\cite{GuiltyByAssociation_Khalil2016}, Kopis~\cite{Kopis_Antonakakis2011}, Notos~\cite{Notos_Antonakakis2010}
\\ \cline{2-3}

&
\textit{2. DNS Dataset Dependent vs.\newline\tabindent DNS Dataset Independent}\newline
\tabindent a) Dependent \newline
\tabindent b) Independent
&
\textbf{2a} (\textit{Examples: Number of IP addresses assigned to a domain, Number of common ASNs shared by a pair of domains}): Hu et al.~\cite{MeasurementAndAnalysisOfGlobalIpUsagePatternsOfFastFluxBotnets_Hu2011}, Konte et al.~\cite{DynamicsOfOnlineScamHostingInfrastructure_Konte2009}, Perdisci et al.~\cite{DetectingMaliciousFluxServiceNetworks_Perdisci2009}, FluxBuster~\cite{FluxBuster_Perdisci2012}, Khalil et al.~\cite{GuiltyByAssociation_Khalil2016} \newline
\textbf{2b} (\textit{Examples: Hit-count of a domain, N-gram distributions of letters in a domain name}): Exposure~\cite{Exposure_Bilge2011,Exposure_Bilge2014}, Notos~\cite{Notos_Antonakakis2010}, Pleiades~\cite{Pleiades_Antonakakis2012}, Marchal et al.~\cite{ProactiveDiscoveryOfPhishingRelatedDomainNames_Marchal2012}
\\ \cline{2-3}

&
\textit{3. Mono Domain vs.\newline\tabindent Multi Domains}\newline
\tabindent a) Mono\newline
\tabindent b) Multi
&
\textbf{3a} (\textit{Examples: Number of countries which host a domain, Number of distinct IP addresses, N-gram distribution of characters in a domain name, Average TTL value}): Stevanovic et al.~\cite{OnTheGroundTruthProblem_Stevanovic2015}, Exposure~\cite{Exposure_Bilge2011,Exposure_Bilge2014}, Kopis~\cite{Kopis_Antonakakis2011}, Fukuda and Heidemann~\cite{DetectingMaliciousActivityWithDnsBackscatter_Fukuda2015}, DomainProfiler~\cite{DomainProfiler_Chiba2016} \newline
\textbf{3b} (\textit{Examples: Related historic domains features, Client sharing ratio between the connected domain names, Number of shared ASN}):  Segugio~\cite{Segugio_Rahbarinia2015, Segugio_Rahbarinia2016}, Khalil et al.~\cite{GuiltyByAssociation_Khalil2016}, Zou et al.~\cite{DetectingMalwareBasedOnDnsGraphMining_Zou2015}, Smash~\cite{Smash_Zhang2015}, Thomas and Mohaisen~\cite{KinderedDomains_Thomas2014}, Notos~\cite{Notos_Antonakakis2010}, GMAD~\cite{GMAD_Lee2014}, Pleiades~\cite{Pleiades_Antonakakis2012}
\\ \hline

Detection \newline Methods
&
\textit{1. Type of Detection Methods}\newline
\tabindent a) Knowledge based\newline
\tabindent b) Machine learning based\newline
\tabindent \tabindent 1) Supervised\newline
\tabindent \tabindent 2) Semi-supervised\newline
\tabindent \tabindent 3) Unsupervised\newline
\tabindent c) Hybrid approaches
&
\textbf{1a} (\textit{Examples: Degree of co-occurrences between known malicious and unknown domains, Similar NXDomain behavior, Distribution of character frequencies in domain names}): Choi et al.~\cite{BotnetDetectionByMonitoringGroupActivitesInDnsTraffic_Choi2007}, Krishnan et al.~\cite{CrossingTheThreashold_Krishnan2013}, Guerid et al.~\cite{PrivacyPreservingDomainFluxBotnetDetection_Guerid2013}, Yadav et al.~\cite{DetectingAlgorithmicallyGeneratedMaliciousDomainNames_Yadav2010,DetectingAlgorithmicallyGeneratedDomainFluxAttacks_Yadav2012}, Holz et al.~\cite{MeasuringAndDetectingFastFluxServiceNetworks_Holz2008} 
\newline 
\textbf{1b1} (\textit{Examples: Naive Bayes, Decision Tree, Random forest, SVM, Neural networks}): FluXOR~\cite{Fluxor_Passerini2008}, Exposure~\cite{Exposure_Bilge2011, Exposure_Bilge2014}, DomainProfiler~\cite{DomainProfiler_Chiba2016}, Stalmans~\cite{FrameworkForDnsBasedDetectionAndMitigation_Stalmans2011},  Woodbridge et al.~\cite{PredictingDgaWithLSTM_Woodbridge2016}, Kopis~\cite{Kopis_Antonakakis2011}, Fukuda and Heidemann~\cite{DetectingMaliciousActivityWithDnsBackscatter_Fukuda2015}, Hu et al.~\cite{MeasurementAndAnalysisOfGlobalIpUsagePatternsOfFastFluxBotnets_Hu2011}, Mentor~\cite{Mentor_Kheir2014} \newline
\textbf{1b2} (\textit{Examples: Cluster-and-label, Belief propagation, Shortest path, Other graph-based approaches}): Huang and Greve~\cite{LargeScaleGraphMiningForWebReputationInference_Huang2015}, DNSRadar~\cite{DnsRadar_Ma2015}, Manadhata et al.~\cite{DetectingMaliciousDomainsViaGraphInference_Manadhata2014}, Mishsky et al.~\cite{TopologyBasedFlowModel_Mishsky2015}, Zou et al.~\cite{DetectingMalwareBasedOnDnsGraphMining_Zou2015}, Khalil et al.~\cite{GuiltyByAssociation_Khalil2016}, GMAD~\cite{TrackingMultipleCCBotnetsByAnalyzingDnsTraffic_Lee2010, GMAD_Lee2014}, Lee and Lee~\cite{TrackingMultipleCCBotnetsByAnalyzingDnsTraffic_Lee2010}, Gao et al.~\cite{EmpiricalReexaminationOfGlobalDnsBehavior_Gao2013, ReexaminingDnsFromAGlobalRecursiveResolverPerspective_Gao2016}, Felegyhazi et al.~\cite{OnThePotentialOfProactiveDomainBlacklisting_Felegyhazi2010} \newline
\textbf{1b3} (\textit{Examples: K-means, X-means, Hierarchical clustering, Agglomerative clustering, Fast unfolding}): Jiang et al.~\cite{IdentifyingSuspiciousActivitiesThroughDnsFailureGraphAnalysis_Jiang2010}, Stevanovic et al.~\cite{OnTheGroundTruthProblem_Stevanovic2015}, BotGAD~\cite{BotGAD_Choi2009,BotGAD_Choi2012}, Thomas and Mohaisen~\cite{KinderedDomains_Thomas2014}, Smash~\cite{Smash_Zhang2015}, Berger and Gansterer~\cite{ModelingDnsAgilityWithDnsMap_Berger2013} \newline
\textbf{1c} (\textit{Examples: combination of machine learning approaches, mix of machine learning and knowledge based techniques}): Perdisci et al.~\cite{DetectingMaliciousFluxServiceNetworks_Perdisci2009}, Oprea et al.~\cite{DetectionOfEarlyStageEnterpriseInfection_Oprea2015}, FluxBuster~\cite{FluxBuster_Perdisci2012}, Pleiades~\cite{Pleiades_Antonakakis2012}, Notos~\cite{Notos_Antonakakis2010}, Segugio~\cite{Segugio_Rahbarinia2015,Segugio_Rahbarinia2016}, Yu et al.~\cite{SemiSupervisedTimeSeriesModeling_Yu2014}
\\ \hline

Outcome
&
\textit{1.  Malicious Behavior Agnostic vs.\newline\tabindent Malicious Behavior Specific}\newline
\tabindent a) Agnostic\newline
\tabindent b) Specific
&
\textbf{1a}: Zou et al.~\cite{DetectingMalwareBasedOnDnsGraphMining_Zou2015}, Manadhata et al.~\cite{DetectingMaliciousDomainsViaGraphInference_Manadhata2014}, Oprea et al.~\cite{DetectionOfEarlyStageEnterpriseInfection_Oprea2015}, GMAD~\cite{GMAD_Lee2014}, Khalil et al.~\cite{GuiltyByAssociation_Khalil2016}, Mishsky et al.~\cite{TopologyBasedFlowModel_Mishsky2015} \newline
\textbf{1b} (\textit{Examples: DGA detection, FFSN detection}): Yadav et al.~\cite{DetectingAlgorithmicallyGeneratedMaliciousDomainNames_Yadav2010, DetectingAlgorithmicallyGeneratedDomainFluxAttacks_Yadav2012}, Haddadi et al.~\cite{MaliciousAutomaticallyGeneratedDomainNameDetection_Haddadi2013}, Grill et al.~\cite{DetectingDgaMalwareUsingNetflow_Grill2015}, Pleiades~\cite{Pleiades_Antonakakis2012}, DeepDGA~\cite{DeepDGA_Anderson2016}, Fu et al.~\cite{StealthyDomainGenerationAlgorithms_Fu2017}
\\ \hline

\end{tabular}
\setcounter{tab_footnote_cnt}{\value{mpfootnote}}
\end{minipage}
\end{table*}

\subsection{Features}
\label{subsec:features}
Feature extraction (a.k.a., feature engineering) is a challenging task which has a big impact on the quality (accuracy and robustness) of the detection approaches. Well-crafted features contribute considerably to the success of an approach, and on the contrary, poor features may ruin even good detection algorithms. On the other hand, even though a feature may have good predictive power leading to a high detection accuracy, if it can be easily forged by an attacker, the robustness of the detection approaches relying on it will be low. Therefore, successful detection approaches must take into consideration a delicate balance of accuracy and robustness when selecting their features.

Very few approaches simply parse Resource Records from DNS traffic and use values from specific fields as they appear. Instead, a multitude of treatments can be applied to these raw values before consuming them for detection purposes (average, standard deviation, max, min, rate, outlier, etc.). Furthermore, an external data, outside the DNS environment, may be used to enrich the initial dataset. Some approaches require to transform the DNS data into a distinct data structure, such as a graph, before using it in their detection methods. For instance, this is the case in the approach proposed by Lee et al.~\cite{GMAD_Lee2014, TrackingMultipleCCBotnetsByAnalyzingDnsTraffic_Lee2010}, where a graph representing the communication sequences of clients with domains is built. The authors call it the Domain Name Travel Graph (DNTG) and use it to identify clusters of related domains that need to be considered by their detection method. In the approach proposed by Oprea et al.~\cite{DetectionOfEarlyStageEnterpriseInfection_Oprea2015}, another type of graph is built representing the association between host IP addresses and queried domains, while in the Khalil et al.~\cite{GuiltyByAssociation_Khalil2016} approach a graph captures the movement of domains in bulks among different ASNs.

The number of individual treatment, enrichment and preprocessing techniques is very large and going through each and every one is out of the scope of this paper. In order to present the state of the art in a systematic way, we distinguish consumed features at a higher level of abstraction. Specifically, we consider the following three dimensions to differentiate features:

\begin{enumerate}
\item Internal vs. Contextual features
\item DNS dataset Dependent vs. Independent features
\item Mono vs. Multi domains features
\end{enumerate}

\subsubsection{Internal vs. Contextual Features}~\\
The distinction between internal and contextual features is quite similar to the one proposed by Perdisci et al.~\cite{DetectingMaliciousFluxServiceNetworks_Perdisci2009} to divide features into \textit{passive} and \textit{active}. According to the authors, \textit{passive} features are the ones ``that can be directly extracted from the information collected by passively monitoring the DNS queries'' from resolvers, while ``\textit{active} features need some additional external information to be computed''. Since we do consider, elsewhere, the possibility to collect the data passively or actively, we felt this terminology could be misleading and therefore, we opt for the different terms, namely internal and contextual, which are described below.

\paragraph{Internal features}
These features can be extracted from DNS Resource Records alone. No external complimentary data source is required. However, they may be and most of the time are transformed before being fed into the detection method. For instance, the ``domain average TTL value'' used in~\cite{Exposure_Bilge2011, Exposure_Bilge2014, DetectingMaliciousFluxServiceNetworks_Perdisci2009, FluxBuster_Perdisci2012, FrameworkForDnsBasedDetectionAndMitigation_Stalmans2011} is an example of this type of features. Additionally, features extracted from domain names, which are popular in DGA detection and attribution (~\cite{Notos_Antonakakis2010, Pleiades_Antonakakis2012, BotGAD_Choi2012, MeasuringAndDetectingFastFluxServiceNetworks_Holz2008, Phoenix_Schiavoni2014, MethodForDetectingDgaBotnet_Tong2016}), belong to this category. Moreover, association-based features popular in graph-based approaches~\cite{DetectingMalwareBasedOnDnsGraphMining_Zou2015, DetectionOfEarlyStageEnterpriseInfection_Oprea2015, GMAD_Lee2014, Segugio_Rahbarinia2015, Segugio_Rahbarinia2016, TrackingMultipleCCBotnetsByAnalyzingDnsTraffic_Lee2010, MethodForIdentifyingCompromisedClients_Stevanovic2017}, are usually built using internal DNS features.

\paragraph{Contextual features}
On the other hand, \textit{contextual} features are built from the combined DNS and external information sources. For instance, to calculate ``the number of ASNs to which the IP addresses of a domain belong to'' (~\cite{DomainProfiler_Chiba2016, FrameworkForDnsBasedDetectionAndMitigation_Stalmans2011, BotGAD_Choi2012, MeasurementAndAnalysisOfGlobalIpUsagePatternsOfFastFluxBotnets_Hu2011}) the information about the IP-AS mapping is required. In other example~\cite{GuiltyByAssociation_Khalil2016}, the authors use similarity score calculated over the number of different AS numbers to assign a weight to domain-domain associations. Zhang et al.~\cite{Smash_Zhang2015} also exploit associations inferred from WHOIS data for domain clustering.

We note that some contextual features require to query resources controlled by attackers. For instance, Prieto et al.~\cite{BotnetDetectionBasedOnDnsRecordsAndActiveProbing_Prieto2011} use domain web presence as one of the features, i.e., every time when a new domain appears in their list they check if a webpage is available for this domain. One more special type of contextual features employs the enrichment using DNS data itself. For instance, Prieto et al.~\cite{BotnetDetectionBasedOnDnsRecordsAndActiveProbing_Prieto2011} checks if a domain has an associated MX record. Hence, the usage of such type of features may warn an attacker that the domain is under scrutiny. However, it is not always required to interact actively with the domains. Such a data sometimes can be obtained from the systems like Thales~\cite{Thales_Kountouras2016}, Censys~\cite{Censys_Durumeric2015} or Shodan~\cite{Shodan}.

Whereas the usage of internal features has a number of benefits, mostly in terms of simplicity, their ability to capture the information that has been shown to be significant to distinguish between good and bad domain names, is limited. For instance, the registration time for a given domain is often a very important feature but it cannot be obtained solely from the DNS data. It has been shown that sometimes attackers register domains in bulk several months before the start of malicious activities~\cite{Predator_Hao2016}. Detection of such registration patterns enables researchers to proactively detect malicious domains as done in~\cite{OnThePotentialOfProactiveDomainBlacklisting_Felegyhazi2010, Predator_Hao2016}. However, that information usually is not available for country code TLDs (ccTLD) because ccTLD registries very rarely offer access to their zone files. Therefore, the existence of a domain can remain unknown until it is queried for the very first time, and at this moment it may be possible (sometimes but not always) to retrieve that information by querying a WHOIS server. This makes the approaches relying on such features inapplicable for a very large amount of domains. Similarly, some other useful enrichment information can be hard to obtain due to limited accessibility, privacy concerns, excessive cost, etc. However, despite all these issues the usage of contextual information allows researchers to extract more meaningful features and hence, provide broader coverage of malicious behavior signals.

\subsubsection{DNS Dataset Dependent vs. Independent features}~\\
Based on our review of the literature, we believe it is important to distinguish between the features that are influenced by specific DNS datasets and those that are independent from the DNS dataset in hand. We call them DNS Dataset Dependent Features (DDD) and DNS Dataset Independent Features (DDI) respectively. The rationales behind these two classes are linked to the validation phase. The performance of an approach solely relying on DDD features is highly influenced by the chosen dataset. Thus, to evaluate the quality of such methods, it is very important to perform cross-dataset validation, using datasets from different places, for different periods, of different sizes, etc. (see Section~\ref{subsec:evaluation_challenges} for more). On the contrary, approaches relying on DDI features are more stable and can be run equally in different environments.

\paragraph{DNS dataset dependent features}
For instance, ``the number of  IP addresses observed as being assigned to a domain'' during the observation period is a DDD feature, because its value depends on the specific dataset~\cite{MeasurementAndAnalysisOfGlobalIpUsagePatternsOfFastFluxBotnets_Hu2011, DynamicsOfOnlineScamHostingInfrastructure_Konte2009, DetectingMaliciousFluxServiceNetworks_Perdisci2009, FluxBuster_Perdisci2012}. Similarly, ``the number of observed common ASNs shared by a pair of domains'' feature used by Khalil et al.~\cite{GuiltyByAssociation_Khalil2016} to build an association between domain names, is also dataset dependent because a graph built using this association hinges on where and how a dataset has been collected.

\paragraph{DNS dataset independent features}
On the other hand, the ``hit-count of a particular domain in popular search engines''~\cite{Exposure_Bilge2011, Exposure_Bilge2014} is a DNS dataset independent feature because it does not depend on what one can see in the DNS dataset chosen. Similarly, the ``n-gram'' distribution of a domain name~\cite{Notos_Antonakakis2010, Pleiades_Antonakakis2012, ProactiveDiscoveryOfPhishingRelatedDomainNames_Marchal2012} is DNS dataset independent since it does not hinge on the chosen dataset.

\subsubsection{Mono vs. Multi Domains Features}
\paragraph{Mono domain features} Mono Domain features are extracted for every single domain. For example, ``the number of countries which host a given domain''~\cite{OnTheGroundTruthProblem_Stevanovic2015, Notos_Antonakakis2010, Kopis_Antonakakis2011, DetectingMaliciousActivityWithDnsBackscatter_Fukuda2015, DomainProfiler_Chiba2016}, is an example of a Mono Domain feature. One of the advantages of using this type of features is that the approaches rely on them can be trained and operate on completely different datasets.

\paragraph{Multi domains features}
Domain association features calculated over a pair of domains, which are used in many graph-based and clustering approaches~\cite{Segugio_Rahbarinia2015, Segugio_Rahbarinia2016, GuiltyByAssociation_Khalil2016, DetectingMalwareBasedOnDnsGraphMining_Zou2015, Smash_Zhang2015, KinderedDomains_Thomas2014}, are examples of Multi domains features and so are the ones used, for instance, in~\cite{Notos_Antonakakis2010, GMAD_Lee2014, Pleiades_Antonakakis2012}. We note that the approaches relying on \textit{Multi Domains} features usually require bigger datasets to work properly. Indeed, an association between two arbitrary domains may be indirect, hence in order to build such an association intermediate domains should be also included into consideration in order for the approach to work properly.

\subsection{Detection Methods}
\label{subsec:methods}
We have identified two main paradigms in the detection methods we are considering. In the first, the method may benefit from some external expertise to figure out how to discriminate between good and bad domains. This expertise is implemented by means of various heuristics and does not use machine learning techniques. Therefore, we call the approaches under this paradigm \textbf{Knowledge Based methods}.
In the second case, whereas the authors may have some examples of malicious and benign domains at their disposal, they have no a priori understanding on how to distinguish between the two. They rely on data driven algorithms to help automatize the discrimination process, therefore, we call the approaches under this paradigm as \textbf{Machine Learning Based methods}.

In general, the approaches belonging to the former category appear earlier than the latter. In early research efforts, through the analysis of data, researchers identified characteristics that allowing to distinguish malicious domains from benign ones. However, with the lapse of time adversaries adapted their behavior causing the degradation of the approaches' detection abilities, what forced researchers to look for more descriptive characteristics. Such races resulted in the situation when the number of characteristics required to be considered in one model, became unmanageable, pushing researchers to look towards the \textbf{Machine Learning Based methods} able to automatically derive knowledge from high-dimensional data.

With a further development of the field, researchers started to employ stacking of methods. In order to produce a list of malicious domains, these methods involve several steps when the output of one method is passed as an input to the following one. So as these techniques employ different detection methods including machine learning and knowledge based, we call them as \textbf{Hybrid approaches}.

\subsubsection{Knowledge Based Approaches}~\\
To detect domains involved in malicious activities, knowledge based approaches rely on expert insights. Such insights can be obtained through measurement studies, which explore anomalies relevant to malicious domain activities. There is a number of such studies in the literature~\cite{DnsMeasurementsAtRootServer_Brownlee2001, PassiveMonitoringOfDnsAnomalies_Zdrnja2007, DayAtTheRootOfTheInternet_Castro2008, AnalyzingDnsActivitiesOfBotProcesses_Morales2009, BayesianBotDetectionBasedOnDnsTrafficSimilarity_VillamarinSalomon2009, BotnetDetectionByMonitoringGroupActivitesInDnsTraffic_Choi2007, BotnetDetectionBasedOnDnsRecordsAndActiveProbing_Prieto2011, CrossingTheThreashold_Krishnan2013,DetectingDgaMalwareUsingNetflow_Grill2015, ExtendingBlackDomainNameListByUsingCoOccurrenceRelationBetweenDnsQueries_Sato2010, PrivacyPreservingDomainFluxBotnetDetection_Guerid2013}. For instance, Sato et al.~\cite{ExtendingBlackDomainNameListByUsingCoOccurrenceRelationBetweenDnsQueries_Sato2010} observed malicious domains belonging to one malware family tend to be queried simultaneously. Hence, by measuring a degree of co-occurrences between known malicious and unknown domains and by comparing the result with some threshold, it is possible to detect new malicious domains. Hyunsang Choi exploited the same observation in his works~\cite{BotnetDetectionByMonitoringGroupActivitesInDnsTraffic_Choi2007,BotGAD_Choi2009,BotGAD_Choi2012}. Krishnan et al.~\cite{CrossingTheThreashold_Krishnan2013} and Guerid et al.~\cite{PrivacyPreservingDomainFluxBotnetDetection_Guerid2013} observed the communities of bots in a network tend to exhibit similar patterns in terms of DNS queries that can not be resolved by the DNS infrastructure.

Unfortunately, this family of approaches have limitations. Experts can intentionally or most often unintentionally be biased. For instance, Grill et al.~\cite{DetectingDgaMalwareUsingNetflow_Grill2015} built their approach on the observation that the DGA malware makes a lot of DNS resolutions in order to find the right domain to communicate with. Therefore, for hosts infected with this type of malware the amount of DNS resolutions is larger than the amount of subsequent communications. Comparing the ratio between them with a manually set threshold allowed the authors to detect hosts infected the malware. However, modern browsers try to predict users' Internet behavior and resolve ahead of time some domains, even if they are never queried. Hence, in such a scenario, if the threshold is not adjusted automatically, the approach will generate false positives since such behavior was unknown to the experts at the time of analysis. Furthermore, experts usually are not good at analyzing high-dimensional data because for a human being it is difficult to grasp all the correlations and dependencies between features extracted from the data.

\subsubsection{Machine Learning Based Approaches}~\\
The majority of the methods developed to detect malicious domains are data-driven with machine learning algorithms at their core~\cite{OnTheGroundTruthProblem_Stevanovic2015}. Generally, machine learning algorithms allow computers to learn on data without being explicitly programmed~\cite{SomeStudiesInMachineLearning_Samuel1959,MachineLearningBook_Mitchell1997}. Depending on what data is used for learning, existing machine learning techniques can be generally divided into three subcategories:

\begin{itemize}
  \item Supervised learning
  \item Semi-supervised learning
  \item Unsupervised learning
\end{itemize}

\paragraph{Supervised learning algorithms} These algorithms require the complete training set to be labeled, i.e., every feature vector corresponding to a sample of data must be associated with a label representing a class this sample belongs to. With respect to the topic of our paper, this means every domain name in the training set must be explicitly labeled as either malicious or benign. However, considering the amount of domains typically observed during the training period of experiments, it is almost impossible to label all of them correctly. Therefore, usually in case of supervised learning approaches the training data set is trimmed to contain only those labeled with high confidence. Interested readers may refer to~\cite{SupervisedMachineLearning_Kotsiantis2007} for a review of supervised learning algorithms. Supervised machine learning approaches such as~\cite{Fluxor_Passerini2008, FrameworkForDnsBasedDetectionAndMitigation_Stalmans2011, DomainProfiler_Chiba2016, ExecScent_Nelms2013, Exposure_Bilge2011, Exposure_Bilge2014, Kopis_Antonakakis2011, Mentor_Kheir2014, DetectingMaliciousActivityWithDnsBackscatter_Fukuda2015, MeasurementAndAnalysisOfGlobalIpUsagePatternsOfFastFluxBotnets_Hu2011} are quite popular in this area due to their simplicity, automatic selection of the most relevant features and effectiveness. Indeed, researchers relying on such approaches only need to extract features from raw data and train a classifier on a labeled dataset. Application of the trained classifier to new data is straightforward. For example, DomainProfiler~\cite{DomainProfiler_Chiba2016} uses 55 features extracted considering related IP addresses and domain names. The Random Forest algorithm is applied to discover abused domains. Antonakakis et al.~\cite{Kopis_Antonakakis2011} also employs Random Forest. However, in this work the features are extracted from the passive DNS data of authoritative name servers. 

Unfortunately, supervised learning approaches have several drawbacks. First, they require a labeled dataset to train. It is not easy to obtain complete and fully correct dataset because of the fickle nature of DNS and blacklist data. As discussed in Section~\ref{subsec:sources_ground_truth}, manual labeling is time-consuming and does not result in extensive training datasets. Automatic labeling using information from different white- and blacklists likewise is prone to incorrect data inclusion~\cite{PaintItBlack_Kuhrer2014, OnTheGroundTruthProblem_Stevanovic2015, CanDnsBasedBlacklistsKeepUpWithBots_Ramachandran2006, ShadesOfGrey_Sinha2008, EmpiricalAnalysisOfPhishingBlacklists_Sheng2009, EmpiricalResearchOfIpBlacklists_Dietrich2009, EmpiricalAnalysisOfMalwareBlacklists_Kuhrer2012}. Second, supervised learning approaches are more vulnerable to overfitting to a particular dataset. If the labeled dataset is biased, this may unintentionally cause a classifier to learn incorrect distributions of the feature variables. Moreover, in a real feed of DNS data only a portion of domains can be assigned with labels. In practice, the vast majority of samples are not labeled and thus, can not participate in the process of classifier learning making training dataset inconsistent.

\paragraph{Semi-supervised learning algorithms} The semi-supervised learning algorithms~\cite{SemiSupervisedLearningLiteratureSurvey_Zhu2005,SemiSupervisedLearningBook_Chapelle2010} have been proposed to overcome such limitations. They learn both from labeled and unlabeled data. The unlabeled data helps a machine learning algorithm to modify or reprioritize hypothesis obtained from a labeled dataset~\cite{SemiSupervisedLearningLiteratureSurvey_Zhu2005}. Yet, the adoption of such algorithms often is quite challenging and requires more effort from researchers. We refer to~\cite{SemiSupervisedLearningLiteratureSurvey_Zhu2005} and~\cite{SemiSupervisedLearningBook_Chapelle2010} for more information about semi-supervised learning algorithms. Graph-based inference methods are among the most popular approaches under this category~\cite{DetectingMalwareBasedOnDnsGraphMining_Zou2015, DetectingMaliciousDomainsViaGraphInference_Manadhata2014, DetectionOfEarlyStageEnterpriseInfection_Oprea2015, GMAD_Lee2014, GuiltyByAssociation_Khalil2016, TopologyBasedFlowModel_Mishsky2015, TrackingMultipleCCBotnetsByAnalyzingDnsTraffic_Lee2010, LargeScaleGraphMiningForWebReputationInference_Huang2015}. For instance, Manadhata et al.~\cite{DetectingMaliciousDomainsViaGraphInference_Manadhata2014} detected malicious domains applying the belief propagation algorithm to a host-domain graph extracted from the enterprise HTTP proxy logs\footnote{We consider this work because the paper assures the same algorithm can be applied to DNS data.}. Assuming that malicious hosts more likely communicate with malware domains, while benign hosts may only occasionally query malicious domains, and having a feed of initially malicious and benign domains, the authors, using the belief propagation approach were able to assess the marginal probability of unknown domains in the graph to be malicious. In~\cite{DetectingMalwareBasedOnDnsGraphMining_Zou2015}, the authors predicted malicious hosts and domains applying their method on two types of graphs. The first, Domain Query Response Graph (DQRG), is built using the information from DNS query-response pairs: clients' IP addresses are connected with the queried domain names which on their turn are associated with the returned domains' IP addresses. The second, Passive DNS Graph (PDG), is built using domain names, their canonical connections and corresponding IP addresses (CNAME and A resource records) extracted from passive DNS data. Then, belief propogation was applied on these graphs. Contrary to~\cite{DetectingMaliciousDomainsViaGraphInference_Manadhata2014}, where all benign domains' initial score has the same value, Zou et al.~\cite{DetectingMalwareBasedOnDnsGraphMining_Zou2015} assigned the value based on their rank in the Alexa top $K$ list. Mishsky et al.~\cite{TopologyBasedFlowModel_Mishsky2015} applied the Flow algorithm on a domain-IP graph. However, this graph includes, besides weighted domain-IP edges commonly used in this area, also domain-domain and IP-IP edges that represent ``tell me who your friends are and I will tell you who you are'' relation.

The Cluster-and-Label semi-supervised learning technique is also widely used~\cite{EmpiricalReexaminationOfGlobalDnsBehavior_Gao2013, ReexaminingDnsFromAGlobalRecursiveResolverPerspective_Gao2016, TrackingMultipleCCBotnetsByAnalyzingDnsTraffic_Lee2010, GMAD_Lee2014, OnThePotentialOfProactiveDomainBlacklisting_Felegyhazi2010} in the area. Gao et al.~\cite{EmpiricalReexaminationOfGlobalDnsBehavior_Gao2013, ReexaminingDnsFromAGlobalRecursiveResolverPerspective_Gao2016} proposed an approach to detect malicious domains through clustering based on co-occurrence patterns. Clearly, queries to DNS system from the same malicious agents do frequently co-occur, e.g., when a bot tries to resolve algorithmically generated domain names in order to find the IP address of a master. In this case, the same domain names will frequently pop-up together in DNS resolver logs. The authors exploited this observation in the following way. At first, they performed coarse-grained clustering of the traffic. They selected a time window and for every anchor domain (malicious domain from a labeled dataset) measured how often it co-occurs with other domains within the selected time window. They calculated two metrics: terms frequency that shows how often other domain names are queried together with the anchor domain, and inverse document frequency showing how rare other domains are met across all the windows. Using the predefined thresholds for both metrics, the authors selected coarse-grained clusters associated with every anchor domain. Further, to perform fine-grained clustering, every domain is assigned with a bit vector whose length is equal to the number of times anchor domains are met during the observation period. A bit in this vector is set if the query to the domain happens within a small time window with the query to the anchor one. Later, these vectors are clustered using X-means to select fine-grained clusters. Jehyun Lee and Heejo Lee proposed a new approach to build a graph representing a sequence of client-domain communications, which they called \textit{Domain Name Travel Graph (DNTG)}~\cite{GMAD_Lee2014}. A node in this directed graph represents a domain, while an edge is added between two nodes if the corresponding domains have been queried sequentially by the same client. The weight of an edge grows with the increase of the number of transitions between those domains, while the direction of an edge shows the order of the transition. An edge is also associated with the client sharing ratio score that represents Jaccard similarity of the sets of the clients queried the domains. After the graph is built, it is clustered using the values assigned to the edges and some predefined thresholds. Then, the authors mark all the domains in the clusters containing blacklisted domains as malicious.

At the same time, this type of algorithms is not a silver bullet in the case of limited ground truth. The usage of unlabeled data does not always help, hence, researchers must put additional efforts in the validation of the proposed methods. Also, the problems related to obtaining a correctly labeled dataset are relevant here as well.

\paragraph{Unsupervised learning algorithms} The unsupervised learning methods~\cite{IdentifyingSuspiciousActivitiesThroughDnsFailureGraphAnalysis_Jiang2010, ModelingDnsAgilityWithDnsMap_Berger2013, KinderedDomains_Thomas2014, Smash_Zhang2015, BotGAD_Choi2009, BotGAD_Choi2012} have been introduced not only to eliminate the dependence on labeled datasets. Unsupervised learning approaches, aka clustering techniques~\cite{DataClusteringReview_Jain1999}, automatically divide domains into clusters using only the internal properties of data. In theory, by careful selection of the features which exibits a completely different behavior for malicious and benign domains, it is possible to enable clustering algorithms to divide the provided samples into two clusters. Then, a researcher decides what cluster contains malicious and benign domains~\cite{BotGAD_Choi2009,BotGAD_Choi2012,OnTheGroundTruthProblem_Stevanovic2015}. However, some approaches, e.g.,~\cite{KinderedDomains_Thomas2014, Smash_Zhang2015}, do not follow this path and make a step further. They group domains across several dimensions related to different malicious behaviors, and then select the clusters of malicious domains by correlating the identified groups among each other. 

Although such approaches have a clear benefit in terms of independence over the labeled data, they are not very common in the literature. We believe this is mainly due to the fact that these techniques are the most difficult to design. Additionally, given that labeled datasets usually exist in this area (althogh neither complete nor fully correct), researchers prefer to explore supervised and semi-supervised methods which are easier to employ.

\subsubsection{Hybrid Approaches}~\\
Despite the fact that a single detection algorithm can be categorized according to the provided classification, the majority of the existing real-world approaches are hybrid and employ several algorithms of different types to produce a result. This can be a combination of machine learning techniques~\cite{DetectingMaliciousFluxServiceNetworks_Perdisci2009, Notos_Antonakakis2010, FluxBuster_Perdisci2012, Pleiades_Antonakakis2012, DetectionOfEarlyStageEnterpriseInfection_Oprea2015}. For instance, such approach is used in the Notos system~\cite{Notos_Antonakakis2010}. It trains 5 meta-classifiers during the first stage to evaluate the closeness of a domain to the predefined group of domains (Popular, Common, Akamai, CDN and Dynamic DNS) using a supervised learning technique. Then the calculated closeness scores are used as features for the second-stage supervised learning algorithm. Oprea et al.~\cite{DetectionOfEarlyStageEnterpriseInfection_Oprea2015} combines a semi-supervised method (belief propagation) with a supervised learning algorithm (linear regression). A mix of machine learning and knowledge based methods is also used in the area~\cite{Segugio_Rahbarinia2015, Segugio_Rahbarinia2016, SemiSupervisedTimeSeriesModeling_Yu2014}. For instance, the Segugio system~\cite{Segugio_Rahbarinia2015, Segugio_Rahbarinia2016} combines graph-based prefiltering with supervised machine learning. It works in the following way. At first, the system, using the DNS data collected before a recursive DNS resolver, builds a host-domain graph. Given a set of benign and malicious domains, and some heuristics, it performs filtering of this graph. It marks the known domain nodes as benign and malicious respectively leaving the rest as unknown. Similarly, the system labels host nodes as malicious if they query one of the malicious domains, and benign that resolve only the benign domains. All other machines are marked as unknown. After this, the system performs pruning of the graph removing: 1) machines querying 5 domains or less; 2) proxy hosts (machines quering substantially more domains than other machines); 3) domains that are queried by only one machine; 4) very popular domains (domains queried by a very large number of machines). Then, every domain node left in the graph is also assigned with the following properties: 1) a set of IP addresses the domain is pointed to during the observation window; 2) how long ago the domain was first queried with respect to the observation time window. Using this information Segugio calculates several features: 1) Machine Behavior Features (the fraction of known infected machines, the fraction of unknown machines, total number of machines); 2) Domain Activity Features (number of days a domain was actively queried during the last 2 weeks, the number of consecutive days a domain was queried); 3) IP Abuse Features (fraction of IPs associated to known malware domains during the selected time window, number of IPs and /24's used by unknown domains during time window). Using these features and supervised machine learning algorithms, the authors predict labels of unknowns.


\subsection{Outcome}
\label{subsec:outcome}
At the end, all we want to know is if a domain is malicious or not. However, the mere term \textit{malicious} can be understood in different ways. For instance, some domains may be involved in spamming or phishing, serving C\&C communications, or simply acting as proxies for other types of campaigns. Among many methods proposed, some are capable to recognize specific types of ``maliciousness'', whereas others are not able to explain why they adjudicate a certain domain is malicious or not. Therefore, in this paper we divide approaches according to the outcome of their operation between those detecting \textit{specific malicious behavior} and those that are \textit{agnostic to malicious behavior}.

\paragraph{Malicious behavior agnostic approaches}
Roughly speaking, \textit{malicious behavior agnostic approaches} do not try to capture particular malicious behavior. Instead, they base their intelligence on different type of \textit{associations} between domains. The approaches of this type~\cite{DetectingMalwareBasedOnDnsGraphMining_Zou2015, DetectingMaliciousDomainsViaGraphInference_Manadhata2014, DetectionOfEarlyStageEnterpriseInfection_Oprea2015, GMAD_Lee2014, GuiltyByAssociation_Khalil2016, TopologyBasedFlowModel_Mishsky2015} will predict maliciousness of domains exploiting connection with the domains constituting the ground truth. Such technique is called sometimes ``guilty by association''~\cite{GuiltyByAssociation_Khalil2016}. If a domain has strong connection with a group of known malicious domains, then most probably, this domain is also involved in malicious activities. For instance, if adult-related domains are used as a ground truth, as a result such approaches will produce the list of domains of the same type, given that these domains make use of the same association. Similarly, if such approaches are fed with spam domains, they will predict domains related to spam activities. At the same time, only few blacklists report malicious domains of particular type, e.g., PhishTank~\cite{PhishTank} or Spamhaus~\cite{Spamhaus}. Moreover, it is usual that the same infrastructure may be used for different malicious activities. Therefore, even if an approach is fed with a ground truth of particular type, the output may include other types of malicious domains. For example, an attacker may use a server with the same IP address that hosts different types of malicious domains. If an approach builds an association between domains according to the common IP addresses, it will establish a connection between these domains.

\paragraph{Malicious behavior specific approaches}
On the contrary, \textit{malicious behavior specific approaches} are built to capture specific features relevant to particular malicious behavior. For instance, there is a number of approaches that specifically try to capture lexical~\cite{DetectingAlgorithmicallyGeneratedMaliciousDomainNames_Yadav2010, DetectingAlgorithmicallyGeneratedDomainFluxAttacks_Yadav2012, MaliciousAutomaticallyGeneratedDomainNameDetection_Haddadi2013} or resolution~\cite{DetectingDgaMalwareUsingNetflow_Grill2015, Pleiades_Antonakakis2012} features suitable for the detection of automatically generated domain names. Some of the approaches extract features detecting multiple malicious activities. So, Bilge et al.~\cite{Exposure_Bilge2011, Exposure_Bilge2014} extract domain name based features which are relevant (although may be not perfect~\cite{DeepDGA_Anderson2016, StealthyDomainGenerationAlgorithms_Fu2017}) for capturing DGAs, and DNS answer-based features (e.g., amount of different IP addresses, TTL values, etc.), which are apt for detection of domains exposing IP fluxing behavior.

\subsection{Challenges}
\label{subsec:approaches_challenges}

\subsubsection{Feature Related Challenges}~\\
Even though the process of finding meaningful features is not easy in other research areas as well, it is especially challenging in the field of malicious domain detection. Features are not only needed to be well crafted to separate benign from malicious domains, but also they have to be resilient to potential manipulation by miscreants. For example, certain DGAs produce easily recognizable names (e.g., ``ccd2.cn'', ``syx4.cn'', ``oif1.cn'', etc.) and one could see this as a powerful feature to identify these malicious domain names. While this is currently true for a very limited number of DGAs, it is trivial for the attacker to render this feature inoperative by simply changing some parameters of the domain generation algorithm. On the other hand, a feature that takes into account the limited capacity of certain resources (e.g., number of public IP addresses) is more robust because it is harder to forge it without impacting negatively the attacker's gain.

Unfortunately, it is not easy to evaluate the robustness of features in a systematic and measurable way. The importance of the problem has been recognized by many researchers, e.g., in~\cite{Smash_Zhang2015, DnsRadar_Ma2015, DetectionOfEarlyStageEnterpriseInfection_Oprea2015, ExecScent_Nelms2013, FluxBuster_Perdisci2012, Psybog_Kwon2014, Kopis_Antonakakis2011, DomainProfiler_Chiba2016, GMAD_Lee2014}. However, up to our knowledge, none of the existing approaches provides a framework that can be used to evaluate quantitatively the robustness of features. Stinson et al.~\cite{TowardsSystematicEvaluationOfTheEvadabilityOfBotnetDetectionMethods_Stinson2008} presents a qualitative high level evaluation of the evadability of some botnet detection approaches. Others, such as Hao et al.~\cite{Predator_Hao2016}, qualitatively discuss the robustness of some of the important features used in their approach. Nevertheless, providing a framework that offers qualitative and quantitative evaluation of the feature robustness remains an open problem that calls for attention from the research community. Such frameworks have to consider simultaneously the features forging complexity and their impact on attack utility. We argue such a framework could be an effective mean against adaptive attackers as it would help researchers and security experts to build detection tools leveraging features whose forging negatively impacts the attackers' benefits.

\subsubsection{Detection Methods Related Challenges}~\\
Even though the effectiveness of a detection method is important and receives due attention in most of the approaches, its performance is somehow overlooked. However, deep performance analysis is as important as effectiveness analysis for practical consideration and real-world deployments. In real-world deployments, the amount and the rate of DNS traffic could be considerably larger than the datasets used in publications. Hence, detection approaches have to be scalable to work in such production systems. Moreover, some approaches require large datasets to train and to tune their detection algorithms. To address this problem, some authors propose to use distributed computing platforms such as Apache Hadoop~\cite{ApacheHadoop} or Apache Giraph~\cite{ApacheGiraph}. Others reduce their dataset sizes by filtering out data elements deemed to be less important. For example, Exposure~\cite{Exposure_Bilge2011, Exposure_Bilge2014} filters out all domains from the Alexa Top 1000 domains~\cite{Alexa} and those that have been queried less than 20 times during a predefined period of time. Unfortunately, such filtering may result in overlooking important sets of domains which could be potentially malicious. In such cases, we need a systematic performance evaluation that takes into account not only the complexity and scalability of a detection method but also the characteristics of the filtering preprocessing steps required for the needed data size reduction.

Next to the performance evaluation challenge, the second one faced by the malicious domain detection methods is related to the latency endured before the detection. Some approaches like~\cite{Exposure_Bilge2011, Exposure_Bilge2014} rely on aggregated data or run in batch mode, and hence, they have to observe a number of DNS requests before being able to make a decision about the malicious status of a domain. However, the delay incurred by such approaches may render them ineffective against domains that serve malicious activities for short periods of time as is the case of domains fluxing. For example, Sheng et al.~\cite{EmpiricalAnalysisOfPhishingBlacklists_Sheng2009} showed that ``63\% of the phishing campaigns lasted for less than two hours''. On the other hand, some approaches leverage real time features (as opposed to aggregates) and can flag domains on the fly. However, non-aggregated features are usually easier to forge comparing to aggregated ones. Both categories of approaches have advantages and limitations, and hence, the optimal selection of one over another is heavily influenced by the deployment environment.

The third challenge is linked with the adaptive nature of the adversaries. They continuously adapt their behavior to evade detection tools, and detection techniques have to regularly retrain and adapt their models to capture such changes. Moreover, this also means that the techniques themselves with the lapse of time become obsolete, making the corresponding approaches no longer possible to use.

The fourth challenge lies in the lack of any systematic way to quantitatively compare and contrast the effectiveness and the efficiency of various domain detection methods. To obtain reliable quantitative results, every approach should be reproducible and measurable. Reproducibility means the results can be regenerated given the same dataset used in initial training, while measurability means the use of quantitative metrics in evaluating effectiveness and performance. Unfortunately, the authors of approaches rarely share datasets and implementation code, possibly due to the privacy, proprietary, and sometimes security related issues, which makes it hard to reproduce the results and considerably complicates the comparison. One way to go over this challenge is to implement tools proposed in these works using information available in public sources such as papers and technical reports. However, the complexity of such tools is usually paramount and the public sources do not contain sufficient and detailed information to provide reasonable implementation of the approach.

\subsubsection{Outcome Related Challenges}~\\
As a result of an algorithm execution, the system predicts if a domain is malicious or not. However, a domain may be malicious in different aspects. For instance, in the obvious case a domain can be defined as malicious because it is used to send spams or to distribute malicious software. Unfortunately, what constitutes a malicious behaviour is not always that well defined. An example is domains hosting adult content. Some approaches, e.g., Predator~\cite{Predator_Hao2016}, consider these domains as malicious because they are often used in spam-related campaigns. Others~\cite{PaintItBlack_Kuhrer2014,DetectingAlgorithmicallyGeneratedMaliciousDomainNames_Yadav2010,Segugio_Rahbarinia2015} consider such domains as benign. At the same time, it is shown that they are often a cause of higher false positive rates, especially if the ground truth contains this type of domains~\cite{ReevaluatingWisdomOfCrowds_Chia2012}. Generally, Wondracek et al.~\cite{IsTheInternetForPorn_Wondracek2010} confirmed adult domains are often used for malware distribution and aggressive marketing, and should not be blindly considered as benign. Hence, researchers should clearly identify in their works which domains are considered as malicious.

\section{Evaluation Methods}
\label{sec:evaluation}
As discussed in the previous section, the majority of DNS-based malicious domain detection approaches leverage machine learning concepts and techniques such as clustering and classification. Therefore, it is natural for them to use the evaluation metrics and strategies that have been developed and used by the machine learning community. However, this area has the unique challenge of adaptive attackers, who continuously change behavior to evade detection. This limits the time and the scope of the validation results and calls for adaptive evaluation strategies. In this section, we present the commonly used evaluation metrics and strategies, and articulate the unique challenges that researchers face when they validate malicious domain detection approaches. Table~\ref{tab:evaluation} provides a short summary of the information considered here.

\begin{table*}[t!]
\caption{Summary of the ``Evaluation Methods'' Section}
\label{tab:evaluation}

\footnotesize
\begin{minipage}{\linewidth}
\setcounter{mpfootnote}{\value{tab_footnote_cnt}}

\centering
\begin{tabular}{|p{10mm}|p{38mm}|p{76mm}|}
\hline
\multicolumn{1}{|c|}{\textbf{Dimension}} & \multicolumn{1}{c|}{\textbf{Categories}} & \multicolumn{1}{c|}{\textbf{Examples and Related Works}} \\ \hline

Metrics
&
\textit{1. Metric Types}\newline
\tabindent a) TPR/Recall\newline
\tabindent b) FPR\newline
\tabindent c) TNR\newline
\tabindent d) FNR\newline
\tabindent e) Precision\newline
\tabindent f) Accuracy\newline
\tabindent g) F1-score\newline
\tabindent h) AUC
&
\textbf{1a}: Haddadi et al.~\cite{AnalyzingStringFormatBasedClassifiersForBotnetDetection_Haddadi2013}, Segugio~\cite{Segugio_Rahbarinia2015,Segugio_Rahbarinia2016}, Krishnan et al.~\cite{CrossingTheThreashold_Krishnan2013}, DomainProfiler~\cite{DomainProfiler_Chiba2016}, Kopis~\cite{Kopis_Antonakakis2011}, Manadhata et al.~\cite{DetectingMaliciousDomainsViaGraphInference_Manadhata2014}, Khalil et al.~\cite{GuiltyByAssociation_Khalil2016}, Zou et al.~\cite{DetectingMalwareBasedOnDnsGraphMining_Zou2015}, Yadav et al.~\cite{DetectingAlgorithmicallyGeneratedMaliciousDomainNames_Yadav2010,DetectingAlgorithmicallyGeneratedDomainFluxAttacks_Yadav2012}, Yadav and Reddy~\cite{WinningWithDnsFailures_Yadav2011}, Notos~\cite{Notos_Antonakakis2010}, Villamarin-Salomon et al.~\cite{BayesianBotDetectionBasedOnDnsTrafficSimilarity_VillamarinSalomon2009} \newline
\textbf{1b}: Haddadi et al.~\cite{AnalyzingStringFormatBasedClassifiersForBotnetDetection_Haddadi2013}, Segugio~\cite{Segugio_Rahbarinia2015,Segugio_Rahbarinia2016}, Krishnan et al.~\cite{CrossingTheThreashold_Krishnan2013}, GMAD~\cite{GMAD_Lee2014}, Kopis~\cite{Kopis_Antonakakis2011}, Manadhata et al.~\cite{DetectingMaliciousDomainsViaGraphInference_Manadhata2014}, Khalil et al.~\cite{GuiltyByAssociation_Khalil2016}, Zou et al.~\cite{DetectingMalwareBasedOnDnsGraphMining_Zou2015}, Yadav et al.~\cite{DetectingAlgorithmicallyGeneratedMaliciousDomainNames_Yadav2010,DetectingAlgorithmicallyGeneratedDomainFluxAttacks_Yadav2012}, Notos~\cite{Notos_Antonakakis2010}, Villamarin-Salomon et al.~\cite{BayesianBotDetectionBasedOnDnsTrafficSimilarity_VillamarinSalomon2009}  \newline
\textbf{1c}: DomainProfiler~\cite{DomainProfiler_Chiba2016}\newline
\textbf{1d}: Oprea et al~\cite{DetectionOfEarlyStageEnterpriseInfection_Oprea2015}, Stevanovic et al.~\cite{OnTheGroundTruthProblem_Stevanovic2015}, Qian et al.~\cite{OnNetworkLevelClustersForSpamDetection_Qian2010} \newline
\textbf{1e}: GMAD~\cite{GMAD_Lee2014}, DomainProfiler~\cite{DomainProfiler_Chiba2016}, Fukuda and Heidemann~\cite{DetectingMaliciousActivityWithDnsBackscatter_Fukuda2015} \newline
\textbf{1f}: Hsu et al.~\cite{FastFluxBotDetectionInRealTime_Hsu2010}, Stevanovic et al.~\cite{OnTheGroundTruthProblem_Stevanovic2015}, Fukuda and Heidemann~\cite{DetectingMaliciousActivityWithDnsBackscatter_Fukuda2015} \newline
\textbf{1g}: DomainProfiler~\cite{DomainProfiler_Chiba2016}, Fukuda and Heidemann~\cite{DetectingMaliciousActivityWithDnsBackscatter_Fukuda2015}, Haddadi et al.~\cite{MaliciousAutomaticallyGeneratedDomainNameDetection_Haddadi2013} \newline
\textbf{1h}: DomainProfiler~\cite{DomainProfiler_Chiba2016}, Manadhata et al.~\cite{DetectingMaliciousDomainsViaGraphInference_Manadhata2014}, Pleiades~\cite{Pleiades_Antonakakis2012}, Huang and Greve~\cite{LargeScaleGraphMiningForWebReputationInference_Huang2015}, Exposure~\cite{Exposure_Bilge2011,Exposure_Bilge2014}, FluxBuster~\cite{FluxBuster_Perdisci2012}
\\ \hline

Evaluation \newline Strategies
&
\textit{1. Evaluation Types}\newline
\tabindent a) Whole dataset\newline
\tabindent b) One round train-test split\newline
\tabindent c) Leave-p-out cross-validation\newline
\tabindent d) K-fold cross-validation\newline
\tabindent e) Cross-networks validation\newline
\tabindent f) Cross-time validation\newline
\tabindent g) Cross-blacklists validation
&
\textbf{1a}: Choi et al.~\cite{BotnetDetectionByMonitoringGroupActivitesInDnsTraffic_Choi2007}, Villamarin-Salomon et al.~\cite{IdentifyingBotnetsUsingAnomalyDetectionTechniques_VillamarinSalomon2008}, Felegyhazi et al.~\cite{OnThePotentialOfProactiveDomainBlacklisting_Felegyhazi2010}, Hu et al.~\cite{MeasurementAndAnalysisOfGlobalIpUsagePatternsOfFastFluxBotnets_Hu2011}, BotGAD~\cite{BotGAD_Choi2009,BotGAD_Choi2012}, Gao et al.~\cite{EmpiricalReexaminationOfGlobalDnsBehavior_Gao2013,ReexaminingDnsFromAGlobalRecursiveResolverPerspective_Gao2016}, ExecScent~\cite{ExecScent_Nelms2013}, Guerid et al.~\cite{PrivacyPreservingDomainFluxBotnetDetection_Guerid2013}, Stevanovic et al.~\cite{OnTheGroundTruthProblem_Stevanovic2015, OnTheGroundTruthProblem_Stevanovic2015}, Smash~\cite{Smash_Zhang2015}, Predator~\cite{Predator_Hao2016}, PsyBoG~\cite{Psybog_Kwon2014} \newline
\textbf{1b}: Lee and Lee~\cite{TrackingMultipleCCBotnetsByAnalyzingDnsTraffic_Lee2010}, Haddadi et al.~\cite{MaliciousAutomaticallyGeneratedDomainNameDetection_Haddadi2013}, Mentor~\cite{Mentor_Kheir2014}, Fukuda et al.~\cite{DetectingMaliciousActivityWithDnsBackscatter_Fukuda2015}, Oprea et al.~\cite{DetectionOfEarlyStageEnterpriseInfection_Oprea2015} \newline
\textbf{1c}: \newline
\textbf{1d}: Fluxor~\cite{Fluxor_Passerini2008}, Holz et al.~\cite{MeasuringAndDetectingFastFluxServiceNetworks_Holz2008}, Notos~\cite{Notos_Antonakakis2010}, Sato et al.~\cite{ExtendingBlackDomainNameListByUsingCoOccurrenceRelationBetweenDnsQueries_Sato2010}, Exposure~\cite{Exposure_Bilge2011,Exposure_Bilge2014}, Kopis~\cite{Kopis_Antonakakis2011}, Pleiades~\cite{Pleiades_Antonakakis2012}, FluxBuster~\cite{FluxBuster_Perdisci2012}, Manadhata et al.~\cite{DetectingMaliciousDomainsViaGraphInference_Manadhata2014}, Zou et al.~\cite{DetectingMalwareBasedOnDnsGraphMining_Zou2015}, DomainProfiler~\cite{DomainProfiler_Chiba2016}, Khalil et al.~\cite{GuiltyByAssociation_Khalil2016}\newline
\textbf{1e}: Segugio~\cite{Segugio_Rahbarinia2015, Segugio_Rahbarinia2016}, DNSRadar~\cite{DnsRadar_Ma2015} \newline
\textbf{1f}: Segugio~\cite{Segugio_Rahbarinia2015, Segugio_Rahbarinia2016}, Yadav et al.~\cite{DetectingAlgorithmicallyGeneratedMaliciousDomainNames_Yadav2010,DetectingAlgorithmicallyGeneratedDomainFluxAttacks_Yadav2012}, Pleiades~\cite{Pleiades_Antonakakis2012}, DomainProfiler~\cite{DomainProfiler_Chiba2016}, Predator~\cite{Predator_Hao2016} \newline
\textbf{1g}: Stalmans and Irwin~\cite{FrameworkForDnsBasedDetectionAndMitigation_Stalmans2011}, Segugio~\cite{Segugio_Rahbarinia2015,Segugio_Rahbarinia2016}, Felegyhazi et al.~\cite{OnThePotentialOfProactiveDomainBlacklisting_Felegyhazi2010}
\\ \hline

\end{tabular}
\setcounter{tab_footnote_cnt}{\value{mpfootnote}}
\end{minipage}
\end{table*}

\subsection{Metrics}
\label{subsec:metrics}
As mentioned earlier, evaluation is tightly coupled with the ground truth. For the purpose of this section, the ground truth consists of a set of domains labeled either as malicious or benign. Let $P$ and $N$ be the number of malicious and benign domains in the test set,  respectively; $TP$ (True Positives) and $TN$ (True Negatives) be the number of correctly identified malicious and benign domains; and $FP$ (False Positives) and $FN$ (False Negatives) be the number of benign domains that have been incorrectly identified as malicious and the number of malicious domains that have been incorrectly identified as benign, respectively. The most commonly used evaluation metrics in this area are:

\begin{itemize}
\item \textbf{True Positive Rate ($TPR$) or Recall}: The ratio of the correctly identified malicious domains to the total number of malicious domains ($TPR = TP / P$); the higher the value is, the better ($ TPR \in [0,1] $).
\item \textbf{False Positive Rate ($FPR$)}: The ratio of the benign domains flagged as malicious to the total number of benign domains ($FPR = FP / N$); the lower the value is, the better ($ FPR \in [0,1] $).
\item \textbf{True Negative Rate ($TNR$)}: The ratio of the correctly identified benign domains to the total number of benign domains ($TNR = TN / N$); the higher the value is, the better ($ TNR \in [0,1] $).
\item \textbf{False Negative Rate ($FNR$)}: The ratio of the malicious domains flagged as benign to the total number of malicious domains ($FNR = FN / P$); the lower the value is, the better ($ FNR \in [0,1] $).
\item \textbf{Precision}: The ratio of the correctly identified malicious domains to the number of all identified malicious domains ($precision = TP / (TP+FP)$); the higher the value is, the better ($ Precision \in [0,1] $).
\item \textbf{Accuracy ($Acc$)}: The ratio of the correctly identified domains to the whole size of the test set ($Acc = (TP+TN) / (P+N)$); the higher the value is, the better ($ Acc \in [0,1] $).
\item \textbf{F1-measure or F1-score}: The harmonic mean of precision and recall ($F_1 = 2*precision*recall / (precision+recall)$); the higher the value is, the better ($ F_1 \in [0,1] $).
\end{itemize}

During the design phase, a detection algorithm is tuned to identify the thresholds that optimize the desired metrics. However, some of these metrics are negatively correlated, i.e., enhancing the value of a desired metric may result in degrading the value of another one. For example, the desire to increase the $TPR$ may result in the undesired increase of the $FPR$. Therefore, detection accuracy is usually assessed based on a discrimination threshold that reflects the dependency of $TPR$ on $FPR$, which is called the \textbf{Receiver Operating Characteristics (ROC)} curve. The ROC graphical representation enables researchers to assess the achieved true positive rate once the value of false positive rate is fixed. Although a ROC curve is a good graphical representation, it cannot serve as a comparative quantitative metric. Therefore, the \textbf{Area Under the ROC Curve (AUC)} has been proposed as a quantitative comparison metric (e.g.,~\cite{DetectingDgaMalwareUsingNetflow_Grill2015}). In general, a system with higher AUC score is better. However, only few approaches report the AUC values what makes it difficult to compare and contrast different methods.

Finally, we note that some approaches use customized metrics to evaluate other important parameters of their system. For example, Khalil et al.~\cite{GuiltyByAssociation_Khalil2016} report the ``expansion'' as the number of newly detected domains for a given number of known malicious domains (seed). Hao et al.~\cite{Predator_Hao2016} defines ``completeness'' as the number of detected domains compared to other blacklists, and ``delay'' as the time it takes blacklists to identify a spammer domain after registration, while Ma et al.~\cite{DnsRadar_Ma2015} use the time-lagging as a metric to evaluate how long it takes other public sources to blacklist a detected domain.

\subsection{Evaluation Strategies}
\label{subsec:strategies}
Malicious domain detection approaches use different evaluation strategies. Most of them borrow strategies from the machine learning community, where cross-validation is one of the most popular technique. In cross-validation, the dataset is split into training and testing parts and multiple rounds are performed using different partitions to reduce variability. The partitioning could be exhaustive as in the case of the \textbf{leave-p-out cross-validation} or non-exhaustive as in the \textbf{k-fold cross-validation}. In leave-p-out cross-validation, $p$ out of the total $n$ observations are used for testing and the remaining observations are used for training. The results are averaged over all possible $p$ combinations out of the $n$ observations, which makes it difficult to apply in practice due to the large number of rounds. Therefore, this strategy is almost not used in the area. The k-fold cross-validation is more practical and hence, more popular (e.g.,~\cite{Fluxor_Passerini2008, MeasuringAndDetectingFastFluxServiceNetworks_Holz2008, Notos_Antonakakis2010, ExtendingBlackDomainNameListByUsingCoOccurrenceRelationBetweenDnsQueries_Sato2010, Exposure_Bilge2011, Kopis_Antonakakis2011, Pleiades_Antonakakis2012, FluxBuster_Perdisci2012, CrossingTheThreashold_Krishnan2013, DetectingMaliciousDomainsViaGraphInference_Manadhata2014, LargeScaleGraphMiningForWebReputationInference_Huang2015, DetectingMalwareBasedOnDnsGraphMining_Zou2015, DomainProfiler_Chiba2016, GuiltyByAssociation_Khalil2016}). According to this strategy the ground truth dataset is divided into $k$ equal parts, where $k-1$ parts are used for training and the remaining part is used for testing. The experiment is repeated $k$ times, changing every time the part used for testing, and the final score is obtained as an average of all the $k$ rounds. Other popular strategies include: (i) \textbf{Validation against the whole dataset} (e.g.,~\cite{BotnetDetectionByMonitoringGroupActivitesInDnsTraffic_Choi2007, IdentifyingBotnetsUsingAnomalyDetectionTechniques_VillamarinSalomon2008, OnThePotentialOfProactiveDomainBlacklisting_Felegyhazi2010,  MeasurementAndAnalysisOfGlobalIpUsagePatternsOfFastFluxBotnets_Hu2011, BotGAD_Choi2012, ReevaluatingWisdomOfCrowds_Chia2012, EmpiricalReexaminationOfGlobalDnsBehavior_Gao2013, ExecScent_Nelms2013, PrivacyPreservingDomainFluxBotnetDetection_Guerid2013, ReexaminingDnsFromAGlobalRecursiveResolverPerspective_Gao2016, OnTheGroundTruthProblem_Stevanovic2015, Smash_Zhang2015, Predator_Hao2016}). According to this strategy, all predictions are verified against the whole ground truth data. This method is popular in unsupervised approaches, where there is no need for a training set. (ii) \textbf{One round train-test split} divides ground truth into two non-overlapping training and testing sets (e.g.,~\cite{TrackingMultipleCCBotnetsByAnalyzingDnsTraffic_Lee2010, MaliciousAutomaticallyGeneratedDomainNameDetection_Haddadi2013, Mentor_Kheir2014, AnalyzingStringFormatBasedClassifiersForBotnetDetection_Haddadi2013, DetectingMaliciousActivityWithDnsBackscatter_Fukuda2015, DetectionOfEarlyStageEnterpriseInfection_Oprea2015}). For example, the authors in~\cite{MaliciousAutomaticallyGeneratedDomainNameDetection_Haddadi2013, AnalyzingStringFormatBasedClassifiersForBotnetDetection_Haddadi2013} use 70\% of the ground truth for training and the rest 30\% for testing.

Although these validation strategies provide quite reliable results from the machine learning community's point of view, they have issues when applied to malicious domain detection. For instance, attackers and benign users in different parts of the world may have different behavior and hence, different organizations have different traffic profiles. For example, the traffic in a governmental organization network is different from that of a supplying company. Thus, the model produced from the data in one part of the world may not be suitable for the data produced in other parts of the world. Moreover, attackers usually change their behavior over time to avoid being detected, therefore, testing on time periods closer to the training time period may produce better results. Finally, an approach may be good at detecting domains belonging to one specific botnet, while performing poorly for other malware types. To address these issues, several cross-dataset strategies were proposed and applied in this area: (i) \textbf{Cross-networks validation}, in which the training and testing datasets are separated in space, i.e., training and testing datasets are collected at different locations (e.g.,~\cite{Segugio_Rahbarinia2015, DnsRadar_Ma2015}); (ii) \textbf{Cross-time validation}, in which training and testing datasets are collected at different time periods (e.g.,~\cite{Segugio_Rahbarinia2015, DetectingAlgorithmicallyGeneratedMaliciousDomainNames_Yadav2010, Pleiades_Antonakakis2012,  DetectingAlgorithmicallyGeneratedDomainFluxAttacks_Yadav2012, DomainProfiler_Chiba2016, Predator_Hao2016}); (iii) \textbf{Cross-blacklists}, in which training and testing datasets are collected from different malware blacklists (e.g.,~\cite{FrameworkForDnsBasedDetectionAndMitigation_Stalmans2011, Segugio_Rahbarinia2015}). Ideally, a system should be trained and tested on completely different data separated both in terms of time and space.

\subsection{Challenges}
\label{subsec:evaluation_challenges}
The first challenge malicious domain detection approaches face with, lies in the difficulty of new knowledge validation. The majority of the approaches validates effectiveness only against part of the ground truth, the testing set, which is usually a small subset of the whole dataset. However, most of the approaches do not systematically show how to validate the predicted malicious domains that are not part of the ground truth. A few detection approaches have partially addressed this challenge (e.g.,~\cite{Pleiades_Antonakakis2012, GuiltyByAssociation_Khalil2016, OnTheGroundTruthProblem_Stevanovic2015, Smash_Zhang2015, Predator_Hao2016}) by one or a combination of the following strategies:

\begin{description}
\item [Cross-inspection.] Newly detected malicious domains are checked against sources of intelligence other than those used for the ground truth collection. However, it is clear that no combination of blacklists covers all existing malicious domains, otherwise, the new approach would generate already known data and thus, would be redundant. Hence, if this technique is applied, and the approach identifies new malicious domains it is impossible to validate them.
\item [Manual content inspection.] The content of newly detected domains is manually checked for malicious traces. In addition to being not scalable~\cite{Fluxor_Passerini2008}, manual inspection is not reliable~\cite{AllYourIframesPointToUs_Provos2008}. The cost of manually crawling and investigating the content of the potentially large number of newly detected domains is prohibitive. Therefore, only a small set of randomly selected domains is usually checked, while the content of the rest remains unverified.
\item [Automatic content inspection.] Newly detected domains are fed into tools that perform automatic content scanning. Automatic verification is not always reliable because the traces of automatic tools could be detected by malicious domain owners or the malicious domain could be proxied by look-like-benign domains~\cite{DetectionAndAnalysisOfDriveByDownloadAttacks_Cova2010,HuntingTheRedFoxOnline_Li2014,EscapeFromMonkeyIsland_Kapravelos2011}. Additionally, malware domains may simply not expose their malicious services to the public but rather target only specific visitors.
\item [Cross-time validation.] Newly detected domains are periodically checked after prediction against reputable commercial and public blacklists, or using manual content checking. However, one caveat of this strategy is that malicious owners may completely abandon domains which had been predicted to be malicious or simply have them behaving benignly~\cite{EmpiricalAnalysisOfPhishingBlacklists_Sheng2009}. Indeed, there is significant evidence that some attackers verify the presence of their resources in public blacklists before launching an attack~\cite{SarvdapSpambot}. Another caveat is that this strategy is affected by the dynamic maliciousness status of some domains over time. Domains being malicious at the detection time may become benign later and vice versa. For example, in February 2016, Linux Mint web server was hacked and used to distribute malicious content~\cite{LinuxMintHack_Murdock2016} but later it has been regained and cleaned. The first transition (malicious to benign) negatively impacts true positives, while the second transition (benign to malicious) does the same with true negatives. 
\end{description}

The second challenge is the absence of a publicly available reference dataset. Although there was an attempt to provide such a dataset (see Los Alamos DNS Dataset for APT Infection Discovery Challenge~\cite{LosAlamosDnsData}), this practice was not widespread. Having a publicly available reference test set is an important step towards providing a benchmark to compare the effectiveness of various approaches, and it can help researchers to further advance the area in a more systematic way. The absence of a reference dataset combined with difficulties in sharing code makes it hard to repeat experiments for systematic comparison of different approaches. However, we admit that attackers change behavior over time to avoid detection moving from one network to another and adjusting their attack methods. Therefore, it may be hard, if not impossible, to collect a reference dataset that covers different deployment environments and survives the dynamic behavior of adversaries. Complementary to this issue is the absence of a reference ground truth data. Different approaches use different sources. As discussed before, such sources may target different malicious activities and hence, cover different domains. Additionally, the lack of sharing among different sources could increase the gap. For example, in~\cite{ReevaluatingWisdomOfCrowds_Chia2012}, out of the 296 and 192 malicious sites that SiteAdvisor and Safe Web have identified, only 8 are common. That is, evaluating based on a ground truth collected from one source may differ from that based on a ground truth collected from another.

The third challenging issue is in building a unified approach for metric calculation. A real DNS data usually consists of domains which are not all covered by white- and blacklists. This leaves the treatment of some metrics to the discretion of researchers. For instance, one can consider all domains appearing in blacklists as malicious while treating all others as benign. Other may take into the consideration only labeled part of domains out of the whole DNS dataset performing metrics computation. Such approaches may considerably influence the results. Along the same line, filtering of a dataset also influences the evaluation results. Indeed, filtering of domains applied in some approaches (e.g.,~\cite{IdentifyingSuspiciousActivitiesThroughDnsFailureGraphAnalysis_Jiang2010, DetectingMaliciousFluxServiceNetworks_Perdisci2009, Exposure_Bilge2011, Exposure_Bilge2014, OnTheGroundTruthProblem_Stevanovic2015}) may influence both positive and negative sides. For instance, in~\cite{Exposure_Bilge2011, Exposure_Bilge2014} Bilge et al. filtered out domains ``queried less than 20 times during the entire monitoring period'' (because some aggregated statistics simply do not work if there are less than this amount of queries). However, among these domains there may be a number of malicious ones. Therefore, filtering out these domains will increase the amount of false negatives. Similarly, the detection rate is also impacted once long-lived domains are removed~\cite{Exposure_Bilge2011, Exposure_Bilge2014}.

Last but not least, sometimes the approaches in this area are compared using the accuracy metric. This metric is not reliable in case of imbalanced datasets, i.e., those where the number of samples of one class is considerably higher than that of other. Such datasets are quite common in the area. Indeed, it is easy to find a large amount of benign domains, e.g., by using Alexa Top 1,000,000 domains~\cite{Alexa}, while the number of malicious domains is limited by the ones available in blacklists. Therefore, it is better either to use the metrics insensitive to imbalanced datasets (e.g., AUC or F1-measure) or to balance the sets before measuring accuracy~\cite{DataMiningForImbalancedDatasets_Chawla2005,TheRoleOfBalancedTrainingAndTestingDatasets_Wei2013}. Finally, even though in the majority of works the results are reported using the TPR and FPR scores, these approaches can be barely compared because the TRP and FPR metrics depend on each other. Therefore, in order to compare two methods one of the metrics' values in both approaches should be fixed.   
\section{Conclusion}
\label{sec:conclusion}
DNS data carry rich traces of the Internet activities, and are a powerful resource to fight against malicious domains that are a key platform to a variety of attacks. In this paper, we presented a large body of research efforts on utilizing DNS data to detect malicious domains. Table~\ref{tab:summary} summarizes our systematization scheme and findings. As our survey shows, to design a malicious domain detection scheme, one has to consider the following major questions: (1) data sources (Section~\ref{sec:datasources}): what types of DNS data, ground truth and auxiliary information are available; 
(2) features and data analysis techniques (Section~\ref{sec:approaches}): how to derive features to match intuitions of malicious behaviors, and what types of detection techniques the malicious domain discovery problem can be mapped to; 
(3) evaluation strategies and metrics (Section~\ref{sec:evaluation}): how well standard evaluation methodologies fit the detection problem in a specific application context, whether there is a need for additional evaluation strategies that better capture the operational settings when a detection scheme is deployed in practice, how to evaluate the robustness of a technique given the adaptive nature of attackers, and what metrics to use for these purposes.

\begin{table*}[t!]
\caption{A Summary of the Proposed Categorization Scheme and Challenges}
\label{tab:summary}

\footnotesize
\begin{minipage}{\linewidth}
\setcounter{mpfootnote}{\value{tab_footnote_cnt}}

\centering
\begin{tabular}{|p{14mm}|p{14mm}|p{38mm}|p{58mm}|}
\hline
\multicolumn{1}{|c|}{\textbf{Realm}} & \multicolumn{1}{c|}{\textbf{Component}} & \multicolumn{1}{c|}{\textbf{Dimension}} & \multicolumn{1}{c|}{\textbf{Challenge}} \\ \hline

\multirowcell{3}{Data\\Sources}
&
DNS Data
&
\textit{1. Where are the Data Collected} \newline
\tabindent a) Host-resolver\tabindent\tabindent
\tabindent b) DNS-DNS \newline
\textit{2. How are the Data Collected} \newline
\tabindent a) Active\tabindent\tabindent
\tabindent b) Passive
&
1. Hard to obtain access to DNS data\newline
2. Hard to share data to run comparative tests
\\ \cline{2-4}

&
{Data\newline Enrichment}
&
\textit{1. Type of the Enrichment Data}\newline
\tabindent a) Geo-location\newline
\tabindent b) ASN\newline
\tabindent c) Registration records\newline
\tabindent d) IP/domain black-/whitelists\newline
\tabindent e) Associated resource records\newline
\tabindent f) Network information
&
1. Historical change of the enrichment information\newline
2. Enrichment information management\newline
3. Limited or payable access to the information
\\ \cline{2-4}

&
Ground Truth
&
\textit{1. Type of the Ground Truth}\newline
\tabindent a) Malicious\tabindent\tabindent
\tabindent b) Benign
&
1. Low-quality public blacklists and mixed content\newline
2. Low-quality/non-representative benign domains\newline
3. Inconsistent domain levels and proprietary data\newline
4. Imbalanced datasets
\\ \hline

\multirowcell{3}{Approaches}
&
Features
&
\textit{1. Internal vs. Contextual}\newline
\tabindent a) Internal\tabindent\tabindent
\tabindent b) Contextual\newline
\textit{2. DNS Dataset Dependent vs. \newline\tabindent DNS Dataset Independent}\newline
\tabindent a) Dependent\tabindent\tabindent
\tabindent b) Independent \newline
\textit{3. Mono Domain vs.\newline\tabindent Multi Domains}\newline
\tabindent a) Mono\tabindent\tabindent
\tabindent b) Multi
&
1. Easiness to forge or to manipulate\newline
2. Reproducibility of the results\newline
3. Data access, data sharing
\\ \cline{2-4}

&
Detection Methods
&
\textit{1. Knowledge Based vs.
\newline\tabindent Machine Learning Based}\newline
\tabindent a) Knowledge based\newline
\tabindent b) Machine learning based\newline
\tabindent \tabindent 1) Supervised learning\newline
\tabindent \tabindent 2) Semi-supervised learning\newline
\tabindent \tabindent 3) Unsupervised learning\newline
\tabindent c) Hybrid approaches
&
1. Feature resilience evaluation\newline
2. Performance and overhead\newline
3. Ability to work in real-time\newline
4. Adaptiveness of adversaries\newline
5. Systematic comparison of techniques
\\ \cline{2-4}

&
Outcome
&
\textit{1. Malicious Behavior Agnostic vs.\newline\tabindent Malicious Behavior Specific}\newline
\tabindent a) Agnostic\tabindent\tabindent
\tabindent b) Specific
&
1. Difficulty to compare approaches aiming at different goals \newline
2. No clear definition what domains are malicious
\\ \hline

\multirowcell{3}{Evaluation}
&
Metrics
&
\textit{1. Metric Types}\newline
\tabindent a) TPR/Recall\tabindent\tabindent
\tabindent b) FPR\newline
\tabindent c) TNR\tabindent\tabindent
\tabindent d) FNR\newline
\tabindent e) Precision\tabindent\tabindent
\tabindent f) Accuracy\newline
\tabindent g) F1-score\tabindent\tabindent
\tabindent h) AUC
&
1. Lack of well agreed upon metrics\newline
2. Lack of problem-specific metric
\\ \cline{2-4}

&
Evaluation Strategies
&
\textit{1. Evaluation Types}\newline
\tabindent a) Whole dataset\newline
\tabindent b) One round train-test split\newline
\tabindent c) Leave-p-out cross-validation\newline
\tabindent d) K-fold cross-validation\newline
\tabindent e) Cross-networks validation\newline
\tabindent f) Cross-time validation\newline
\tabindent g) Cross-blacklists validation
&
1. New knowledge validation\newline
2. Lack of public reference datasets\newline
3. Unified metrics calculation
\\ \hline

\end{tabular}
\setcounter{tab_footnote_cnt}{\value{mpfootnote}}
\end{minipage}
\end{table*}

Our analysis identifies several significant challenges that hinder the advances of the field. First, in terms of data availability, we observe that large-scale real DNS data logs are seldom publicly available, and sharing of such information across organizational boundaries often faces legal, privacy-related or bureaucratic obstacles. Also, in terms of ground truth, there is no widely agreed practice in the community how to build ground truth from noisy public intelligence. Second, in terms of features and detection techniques, we have highlighted a number of challenges such as the resilience of the features, the adaptability of algorithms to evading attackers, and the interpretation of the results. Third, in terms of evaluation strategies and metrics, current research lacks established theoretical foundations and systematic empirical frameworks to evaluate the robustness of malicious domain detection schemes.

Providing a deep overview of the area, identifying existing challenges, and sharing our insights obtained doing the research in this field, we hope this survey will facilitate future research and development of methods and applications to fight against attacks leveraging
malicious domains.

\bibliographystyle{ACM-Reference-Format}

\end{document}